\documentclass{ws-ijbc}
\usepackage{ws-rotating}     
\usepackage{placeins}
\usepackage{amsmath}
\usepackage{tikz}
\usepackage{graphicx}
\usepackage{epstopdf}
\begin{document}

\newcommand{\threepartdef}[6]
{
    \left\{
    \begin{array}{lll}
        #1 & \quad \mbox{for } #2 \\
        #3 & \quad \mbox{for } #4 \\
        #5 & \quad \mbox{for } #6
    \end{array}
    \right.
}

\catchline{}{}{}{}{} 

\markboth{J. Cantis\'{a}n et al.}{Transient dynamics of the Lorenz system with a parameter drift}

\title{Transient dynamics of the Lorenz system with a parameter drift}

\author{Julia Cantis\'{a}n}
\address{Nonlinear Dynamics, Chaos and Complex Systems Group, Departamento de F\'{i}sica, Universidad Rey Juan Carlos \\ Tulip\'{a}n s/n, 28933 M\'{o}stoles, Madrid, Spain}

\author{Jes\'{u}s M. Seoane}
\address{Nonlinear Dynamics, Chaos and Complex Systems Group, Departamento de F\'{i}sica, Universidad Rey Juan Carlos \\ Tulip\'{a}n s/n, 28933 M\'{o}stoles, Madrid, Spain}

\author{Miguel A.F. Sanju\'{a}n}
\address{Nonlinear Dynamics, Chaos and Complex Systems Group, Departamento de F\'{i}sica, Universidad Rey Juan Carlos \\ Tulip\'{a}n s/n, 28933 M\'{o}stoles, Madrid, Spain} \address{Department of Applied Informatics, Kaunas University of Technology \\ Studentu 50-415, Kaunas LT-51368, Lithuania}

\maketitle

\begin{history}
\received{(to be inserted by publisher)}
\end{history}
\begin{abstract}
Non-autonomous dynamical systems help us to understand the implications of real systems which are in contact with their environment as it actually occurs in nature. Here, we focus on systems where a parameter changes with time at small but non-negligible rates before settling at a stable value, by using the Lorenz system for illustration. This kind of systems commonly show a long-term transient dynamics previous to a sudden transition to a steady state. This can be explained by the crossing of a bifurcation in the associated frozen-in system. We surprisingly uncover a scaling law relating the duration of the transient to the rate of change of the parameter for a case where a chaotic attractor is involved. Additionally, we analyze the viability of recovering the transient dynamics by reversing the parameter to its original value, as an alternative to the control theory for systems with parameter drifts. We obtain the relationship between the paramater change rate and the number of trajectories that tip back to the initial attractor corresponding to the transient state.

\end{abstract}


\section{Introduction}

The evolution of a system with time is typically divided into two different regimes: the transient and the steady states. The latter corresponds to the asymptotic dynamics: after some time the system settles in this state for an indefinite time unless it is perturbed. This final state might include not only fixed points, but also limit cycles or chaotic attractors. The dynamics before the system settles in any of these attractors is what we call transient dynamics. This type of dynamics is also very rich and includes, as a classic example, decaying oscillations before a fixed point, but also chaotic motion. When the system behaves in a chaotic way for a finite amount of time before reaching an attractor, it is said to present transient chaos.

Traditionally, the study of dynamical systems has focused on the steady state as transients usually last for a short-time scale and it has been assumed that the system's dynamics can be reflected by the asymptotic behavior of models describing these systems. However, some systems present long-lasting transients compared to the scale of the system. Additionally, the relevant time scales may correspond to the transient regime rather than to the asymptotic one. Furthermore, transients may provide an explanation for sudden regime shifts, even without any underlying change in external conditions.

Transient phenomena in autonomous systems has been studied in the past few years in several scientific disciplines. Much of the work has focused on analyzing the factors that make a system prone to long transients, for example, time-delay \cite{Morozov2016}. Evidence of systems presenting relevant long-lasting transients include ecological systems \cite{Hastings2004,Hastings2018,Morozov2020}, but also different models in neuroscience \cite{Rabinovich2006,Rabinovich2008}, power electronics \cite{Warecki2014}, earthquake activity in seismology \cite{Picozzi2019}, and  gravitational waves \cite{Thrane2015}. Furthermore, transient chaos is relevant in a wide variety of systems ranging from optomechanics \cite{Wang2016} to electronic engineering \cite{Bo-Cheng2010}. For a recent review on transient chaos see \cite{Lai2011}

Here, we focus on the study of transient phenomena in non-autonomous dynamical systems. In particular, in systems where one of the parameters varies slowly with time. The system can be mathematically written as
\begin{equation}
\frac{d \boldsymbol{x}}{dt}=F(\boldsymbol{x}, \boldsymbol{p} (\varepsilon t)),
\end{equation}
where $ \varepsilon $ is very small compared to the natural time scale of the system. This type of systems are fundamental to understand the relation of the system with its environment. Real systems are affected by external conditions, which can be reflected as a gradual change in a parameter. Sometimes, the parameter is controllable. One example is a ferromagnet within a low frequency magnetic field. In other cases, it is not controllable, as in the context of climate dynamics.

Time series with long-lasting and physically relevant transients are often found in systems with parameter drift. Because of the duration of transients one may think that the system is at its steady state. However, for a later time, they show a sudden transition to their real steady state.

Due to the fact that the time dependence of the parameter represents a small perturbation, the associated frozen-in system, that is, the system with fixed parameters, provides useful information about the non-autonomous system \cite{Berglund2000}. In fact, the origin of the previously mentioned sudden transition may be found in the crossing of a bifurcation in the associated frozen-in system. This type of bifurcations are called dynamic bifurcations \cite{Benoit1991}. Recently, they have been also called bifurcation-induced tipping points \cite{Ashwin2012}. Either way, they refer to the regime shift that is produced due to the slow passage through a bifurcation.

Dynamic bifurcations were studied when only regular attractors are involved. The case of the supercritical Hopf bifurcation has been studied in \cite{Neishtadt1987, Neishtadt1988, Baer1989}. As a result, it was found that the appearance of oscillations was delayed when the parameter is slowly drifting and that the delay depends on the rate of the drift, what is called the delay effect. However, when chaotic attractors are involved, dynamic bifurcations have been only studied so far for maps. In particular, the case of a Lorenz-type map when the control parameter varies monotonically with time and when it induces a periodic forcing, has been analyzed in \cite{Maslennikov2013, Maslennikov2018} respectively. Here, we aim to broaden the current knowledge on dynamic bifurcations to flows presenting strange attractors.

Another interesting phenomenon occurs to systems with parameter drift when there is multistability. In this case, the parameter drift may cause the system to tip to another state. Sometimes, this happens because the current attractor disappears and it has to tip to any of the rest of the attractors. More surprisingly, this can happen for certain rates above a critical value, even if the first attractor is still stable. In this case, the system cannot track the attractor and it tips to another one with a certain probability. This has been called rate-induced tipping \cite{Ashwin2012}. Besides, the study of the tipping probabilities for a monodimensional system including the passage through a chaotic attractor was shown in \cite{Kaszas2019}, but they do not consider the case when the chaotic attractor coexists with other attractors.

Our goal here is to study the transient phenomena for a multistable system with parameter drift, in particular the Lorenz system. The nature and duration of the transient dynamics and the transition to the steady state can be explained by the delay effect previously mentioned. The bifurcation diagram of the frozen-in system gives us information about the dynamics that has to be corrected by a scaling law that relates the magnitude of the delay with the parameter rate of change.

The paper is organized as follows. We start in Sec.~\ref{Soverdamped} with the description of the frozen-in system, including the basins of attraction for the starting and ending points of the parameter shift. In Sec.~\ref{dynamic hetoclinic}, we let the parameter evolve with time in such a way that the heteroclinic bifurcation is crossed. We uncover an interesting scaling law that predicts the duration of the transient for the non-autonomous system. Also, we show that the presence of a chaotic attractor induces unpredictability, causing a random tipping.

Next, in Sec.~\ref{reversibility}, we study the possibility of reversing the transient dynamics once the transition to the steady state has taken place by reversing the parameter to its original value. For that purpose, higher parameter rates have to be considered and rate-induced tipping is the phenomenon responsible for the transient state recovery. Finally, the main conclusions of this work are discussed.

\section{The Lorenz system} \label{Soverdamped}

We have chosen the Lorenz system \cite{Lorenz1963} described by the Eqs.~\ref{Lorenz}, since it can be considered as a paradigmatic example of a multistable system that presents both chaos and transient chaos. These equations were proposed by Edward Lorenz as a simple model of convection dynamics in the atmosphere (Rayleigh-Bénard convection). These equations also arise in models of lasers \cite{Li1990} and chemical reactions \cite{Poland1993}, among others. The equations of the system read as follows
\begin{equation}
\begin{split}
\dot{x}=-\sigma x + \sigma y ,\\
\dot{y}=r x -y -xz,\\
\dot{z}=-\beta z +xy,
\end{split}
\label{Lorenz}
\end{equation}
where $ \sigma $, $ \beta $ and $ r $ are the system parameters. In the context of convection dynamics, $ \sigma $ is the Prandtl number and is characteristic of the fluid, $ \beta $ depends on the geometry of the container and has no specific name. Finally, $ r $ is the Rayleigh number and it accounts for the temperature gradient. In this context, $ x $ represents the rotation frequency of convection rolls, while $ y $ and $ z $ correspond to variables associated to the temperature field. We fix the classical parameter values as $\sigma=10$ and $ \beta=8/3 $. And we explore the dynamics of the system in terms of the variation of $ r $ by using a bifurcation diagram.

\begin{figure}[h]
    \begin{center}
        \includegraphics[width=0.75\textwidth ]{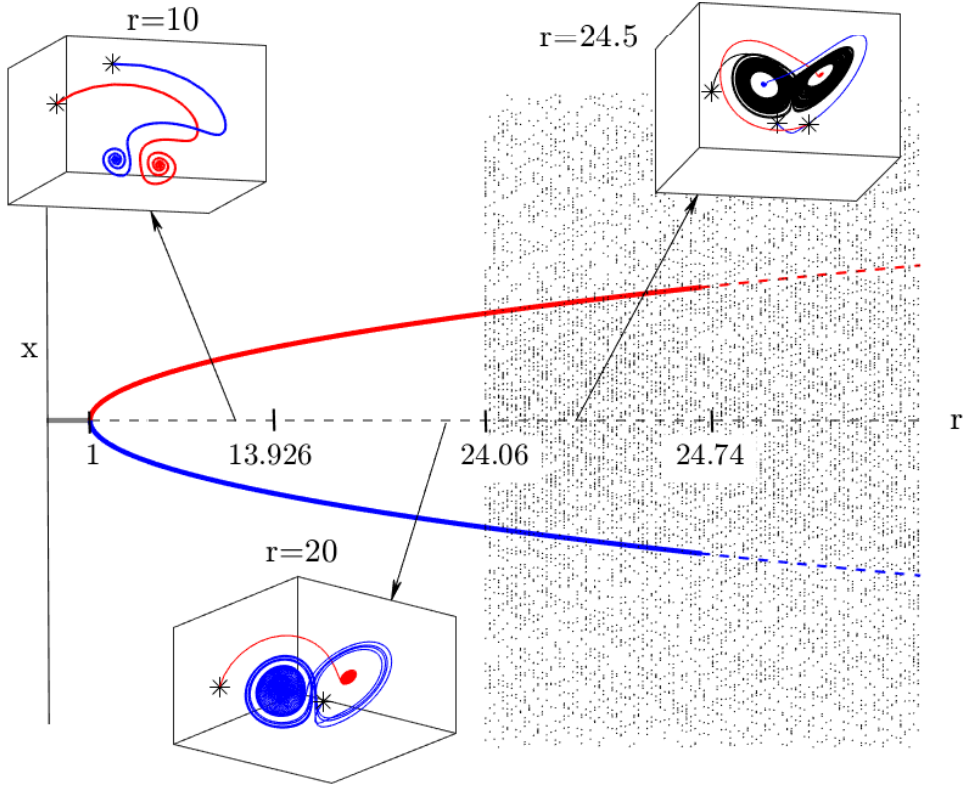} 
    \end{center}
    \caption{\textbf{ Bifurcation diagram for the Lorenz system.} For $ r=1 $ a supercritical pitchfork bifurcation takes place and the origin becomes a saddle point. For $ 1 < r < 24.06 $, two attractors coexist: the upper branch corresponds to $ C^{+} $ and the lower branch corresponds to $ C^{-} $. For this range of $r$ values, past the homoclinic bifurcation (when $ r > 13.926 $), trajectories may be chaotic before settling to one the two fixed point attractors. For $ r > 24.06 $, the chaotic attractor is born in a heteroclinic bifurcation and the three attractors coexist. Finally, for $ r > 24.74 $ a subcritical Hopf bifurcation leaves the chaotic attractor as the global attractor. In the insets, the attractors are represented in phase space for the regions where multistability is present (the stars mark the initial conditions). }
    \label{Bifurcation_diagram}
\end{figure}

Depending on the temperature gradient, i.e., the value of $ r $, convection rolls may exist or not. This can be seen in detail in the bifurcation diagram shown in Fig.~\ref{Bifurcation_diagram}). For $ r < 1 $ the only attractor is $ (0,0,0) $, what means that the fluid remains at rest. At $ r = 1 $, the origin becomes a saddle through a supercritical pitchfork bifurcation, and this instability is reflected in the bifurcation diagram by a discontinuous line for $ r > 1 $. Two symmetrical branches
\begin{equation}
 C^{\pm} =  (\pm \sqrt{b(r-1)}, \pm \sqrt{b(r-1)}, r-1 )
 \label{C+-}
\end{equation}
are created and are stable until $ r=24.74 $. The unstable manifold of the origin, $ W^{s}(\boldsymbol{0}) $, separates their respective basins of attraction. The fixed points attractors $  C^{\pm}  $ correspond to convection rolls with the two possible directions of rotation.

On the other hand, at $ r = 24.06 $, a chaotic attractor is born through an heteroclinic bifurcation which makes the fluid become turbulent. For $ 24.06 < r < 24.74 $, the system is multistable ($ C^{+}, C^{-} $ and the chaotic attractor coexist) and the attractor to which a given trajectory goes to depends on the initial conditions.

Furthermore, at $ r=24.74 $, the fixed points $ C^{\pm} $ lose the stability through a subcritical Hopf bifurcation and the chaotic attractor becomes the global attractor. Another important phenomenon occurs at $ r=13.926 $, when a homoclinic bifurcation takes place. This implies that a chaotic saddle is born, making some trajectories rattle around chaotically for a while before they settle down to $ C^{\pm} $, which is known as transient chaos or preturbulence. This is not reflected in the bifurcation diagram as it only shows the steady state dynamics. In phase space, the chaotic behavior is confined to the vicinity of the chaotic saddle \cite{Tel2006}. The lifetime of these chaotic transients increases with $ r $ until it reaches the value $ 24.06 $ when the lifetime becomes infinite and the chaotic attractor appears.

Also, at $ r=13.926 $ a pair of unstable limit cycles $ \Gamma^{\pm} $ (not represented in the diagram), called homoclinic orbits, are created and last until they are absorbed in the Hopf bifurcation at $ r=24.74 $. For further details about the homoclinic and heteroclinic bifurcations in terms of the organization of the respective two-dimensional manifolds of $ \boldsymbol{0} $, $ C^{\pm} $ and $ \Gamma^{\pm} $ see \cite{Doedel2006}.

Now, we focus on the dynamics before and after $ r=24.06$ in order to explore the effects of a parameter drift when a strange attractor appears/disappears. This is why we present the basins of attraction for $ r=20 $, where transient chaos is present, and $ r=24.5 $, where the chaotic attractor and the fixed point attractors $ C^{\pm} $ coexist.

For this purpose, we distribute $ N=10^5 $ initial conditions uniformly, preserving approximately the same density of points for any area on the sphere, that is, avoiding accumulation of points in the poles. The radius of the sphere is fixed to $ 30 $ for all the simulations in this paper, but similar results are found for other radii. The sphere is likewise centered at the halfway between $ C^{\pm} $: $ (0,0,r-1) $. The basins are computed integrating the trajectories starting on the sphere by a Runge-Kutta algorithm with adaptive step size control. The criterion for convergence to a particular attractor is that the trajectory enters a sphere of radius $ 0.1 $ centered at $ C^{+}/ C^{-} $. For the basin at $ r=24.5 $, we take a sufficiently long integration time (the maximum time for trajectories to arrive to $ C^{\pm} $ is around $ 90 $ and we take $ t_{f}=2000 $ as the final integration time) and we consider that trajectories that do not converge to $ C^{+}/ C^{-} $, converge to the chaotic attractor.

\begin{figure}[h]
    \begin{center}
        \includegraphics[width=0.45\textwidth ]{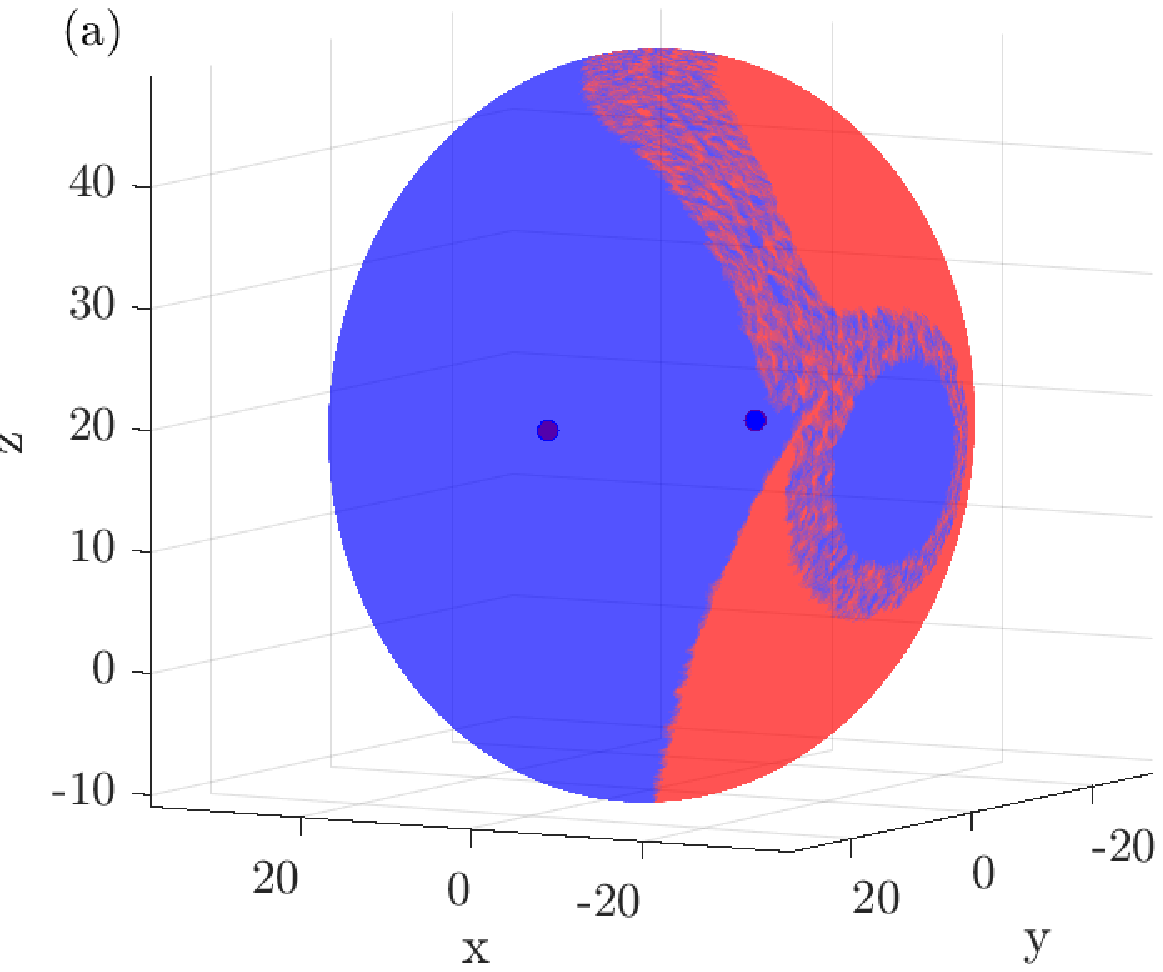}
        \includegraphics[width=0.45\textwidth]{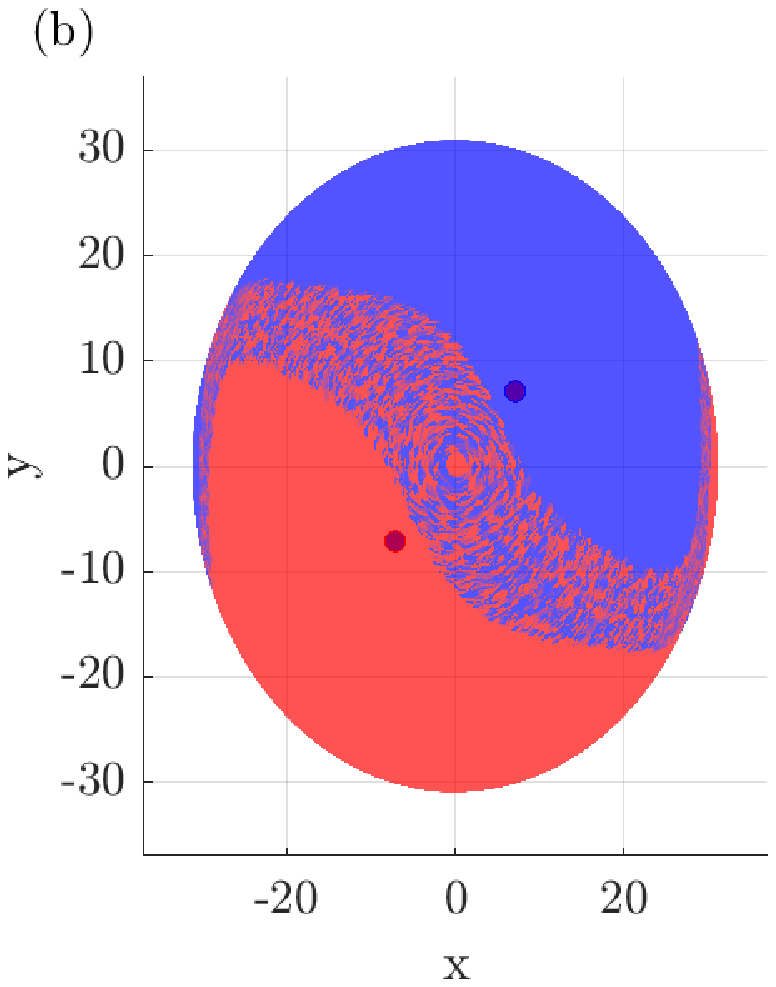}
    \end{center}
    \caption{\textbf{ Basins of attraction for $ r=20 $.} (a) The sphere of initial conditions is represented with the trajectories that end up in the $ C^{+} $ attractor depicted in red and the ones that end up in $ C^{-} $ depicted in blue. The attractors are also represented in the inside as points with matching colors. (b) We present a view in the $(x,y)$ plane. Fractality appears in the basin boundaries for $ z>(r-1)=19 $. }
    \label{basin_r20}
\end{figure}

The basins of attraction for $ r=20 $ are represented in Fig.~\ref{basin_r20}: in red, the initial conditions that end up in $ C^{+} $, and in blue the ones that end up in $ C^{-} $; the attractors are also represented in the inside as points with matching colors. The sphere is divided in two, following approximately the symmetry plane $ x+y=0 $. More precisely, the basin boundary is defined by the stable manifold of the origin $ W^{s}(\boldsymbol{0}) $. It is important to notice that the blue face is nearer the red attractor and vice versa. There is also a circular blue region that is immersed in the red region, and symmetrically there is a circular red region in the backwards of Fig.~\ref{basin_r20}(a). Finally, the basin boundary is smooth for $ z<(r-1)=19 $ (downside) and fractal for $  z>(r-1)=19 $ (upside) and around the immersed circle. This fractality is a trace of chaos.

\begin{figure}[h]
    \begin{center}
        \includegraphics[width=0.45\textwidth ]{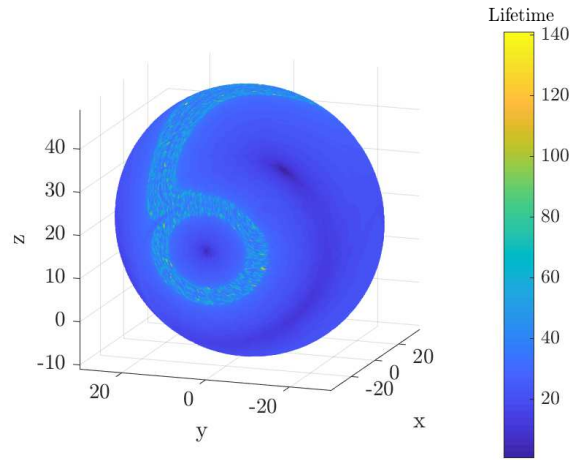}
    \end{center}
    \caption{\textbf{Time to reach the attractors for $ r=20 $.} The numerically computed lifetime of the transients is represented showing a wide range of lifetimes, denoted in the color bar, with higher values on the corresponding fractal regions from Fig.~\ref{basin_r20}. This corresponds to the trajectories presenting transient chaos.}
    \label{lifetime_r20}
\end{figure}

The time that the trajectory needs to reach the attractors is depicted in Fig.~\ref{lifetime_r20}. The lifetime is increased for higher values of the temperature gradient, if $ r<24.06 $. Comparing this figure with Fig.~\ref{basin_r20}, it can be seen that the fractal areas correspond to initial conditions that take the longest time to reach the attractors. The explanation for this phenomenon is related to the stable and unstable manifolds of the chaotic saddle. Points exactly on the stable manifold necessarily reach the chaotic saddle and never leave it (these points are exceptional and are not shown), and points initially far from the stable manifold escape the chaotic saddle quickly (not showing transient chaos), while points near the stable manifold have a longer lifetime. The closer to the stable manifold, the longer the lifetime of the transients is. Thus, the structure seen in light blue in Fig.~\ref{lifetime_r20} is due the cut of the stable manifold of the chaotic saddle with the sphere of initial conditions.

\begin{figure}[h]
    \begin{center}
        \includegraphics[width=.3\textwidth ]{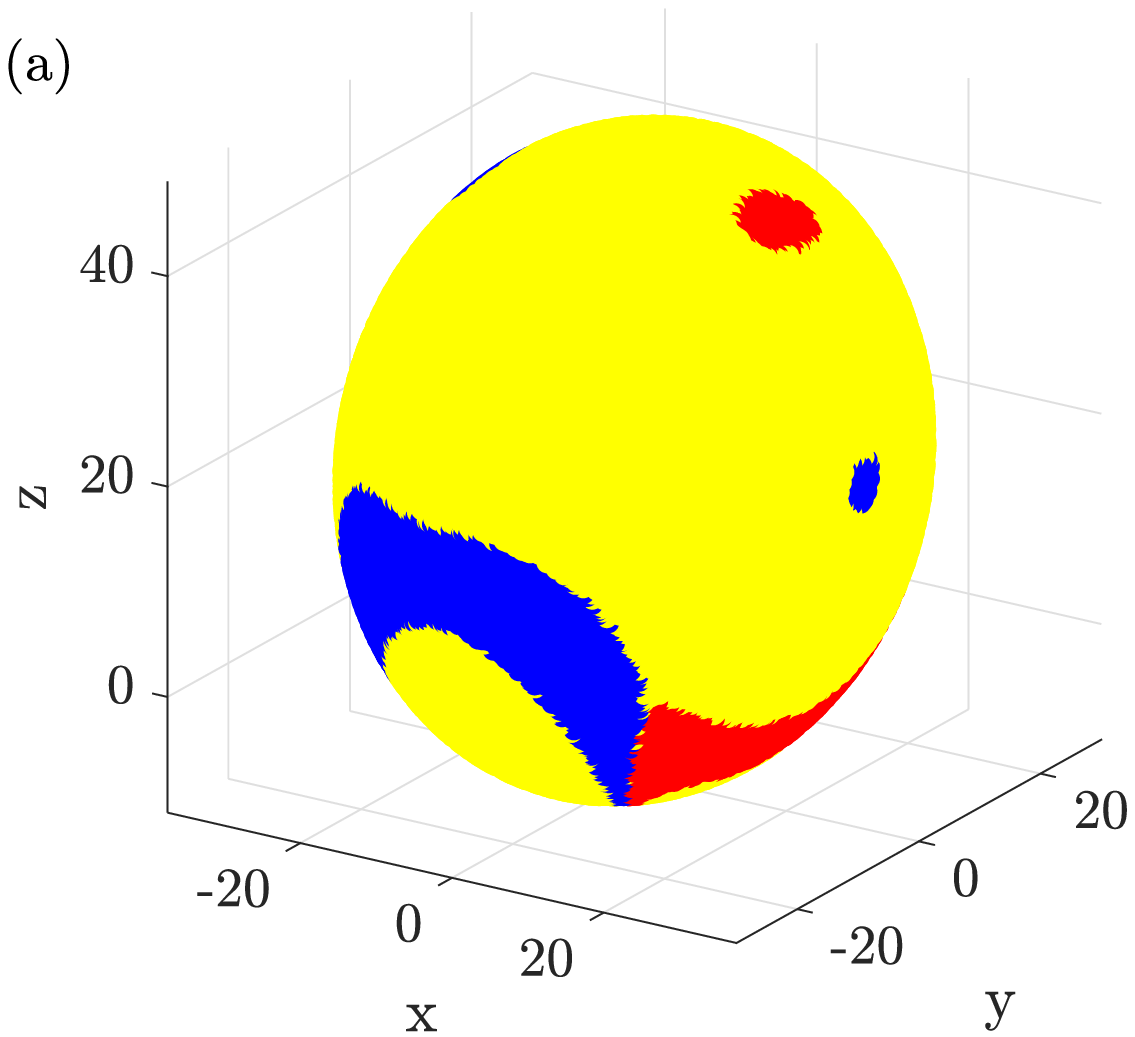}
            \includegraphics[width=.3\textwidth ]{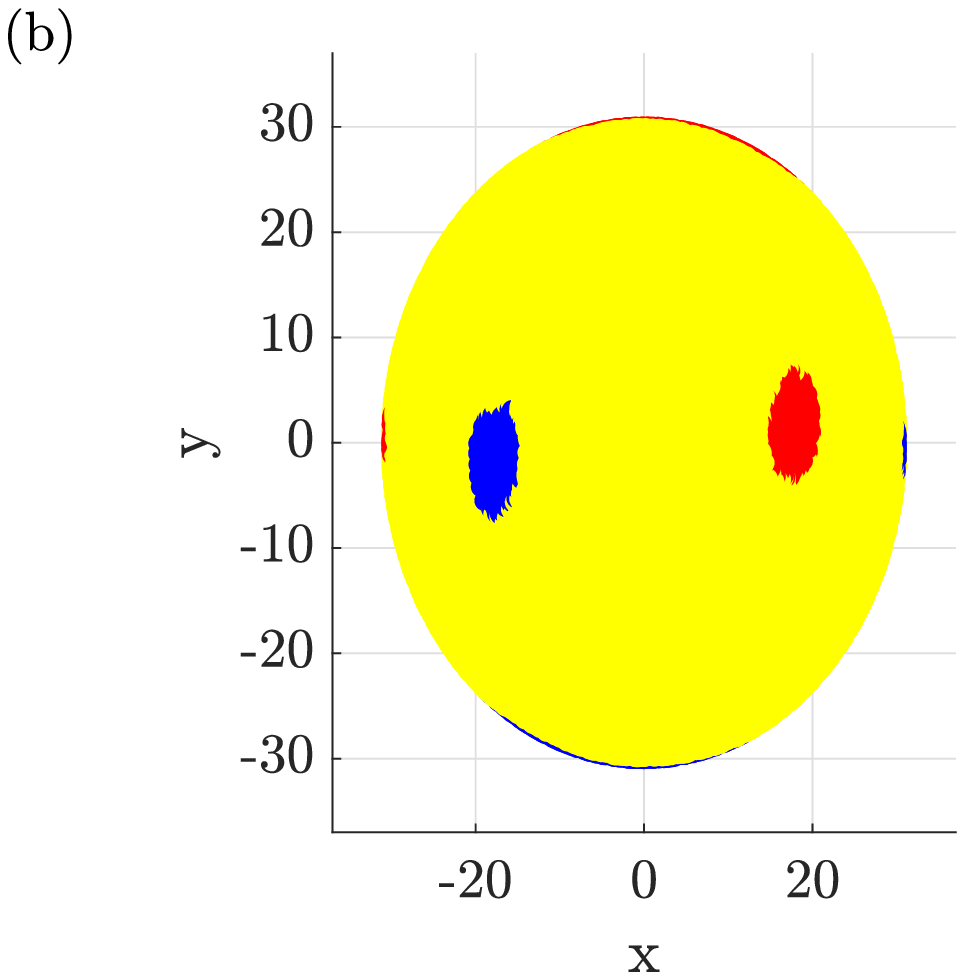}
                \includegraphics[width=.3\textwidth ]{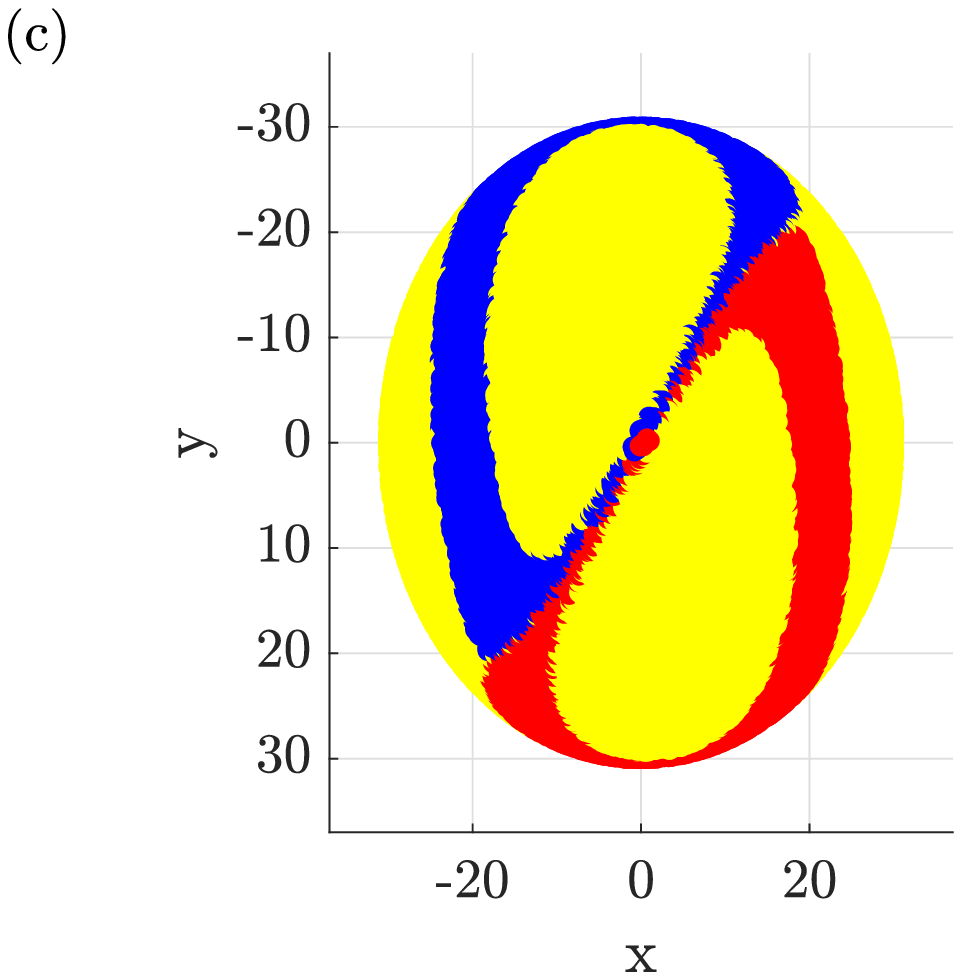}

    \end{center}
    \caption{\textbf{Basins of attraction for $ r=24.5 $.} (a) The sphere of initial conditions is represented with the same color code as before: red accounts for $ C^{+} $ and blue for $ C^{-} $. Yellow regions account for trajectories that head to the chaotic attractor. The (b) and (c) panels show the $(x,y)$ plane for $ z>19 $ (view from above) and $ z<19 $ (view from the downside) respectively. }
    \label{basin_r245}
\end{figure}

The basins of attraction for $ r=24.5 $ are represented in Fig.~\ref{basin_r245}. Following the same color code as before, red/blue regions correspond to initial conditions that end up in $
C^{+}/C^{-} $. The initial conditions that end up in the chaotic attractor are depicted in yellow. For the $ 10^{5} $ trajectories, $ 10348 $ correspond to the red basin, $ 10130 $ to the blue basin and $ 79522 $ to the yellow basin. Thus, approximately the $ 20 \% $ of the trajectories go to $ C^{+}/C^{-} $. This time, the structure for the red and blue basins consists on two loops for $ z<19 $ that overlap in the plane $ x+y=0 $ and four circular regions for $ z>19 $ (one of each color on top and another two smaller ones on the sides).

\begin{figure}[h]
    \begin{center}
        \includegraphics[width=0.45\textwidth ]{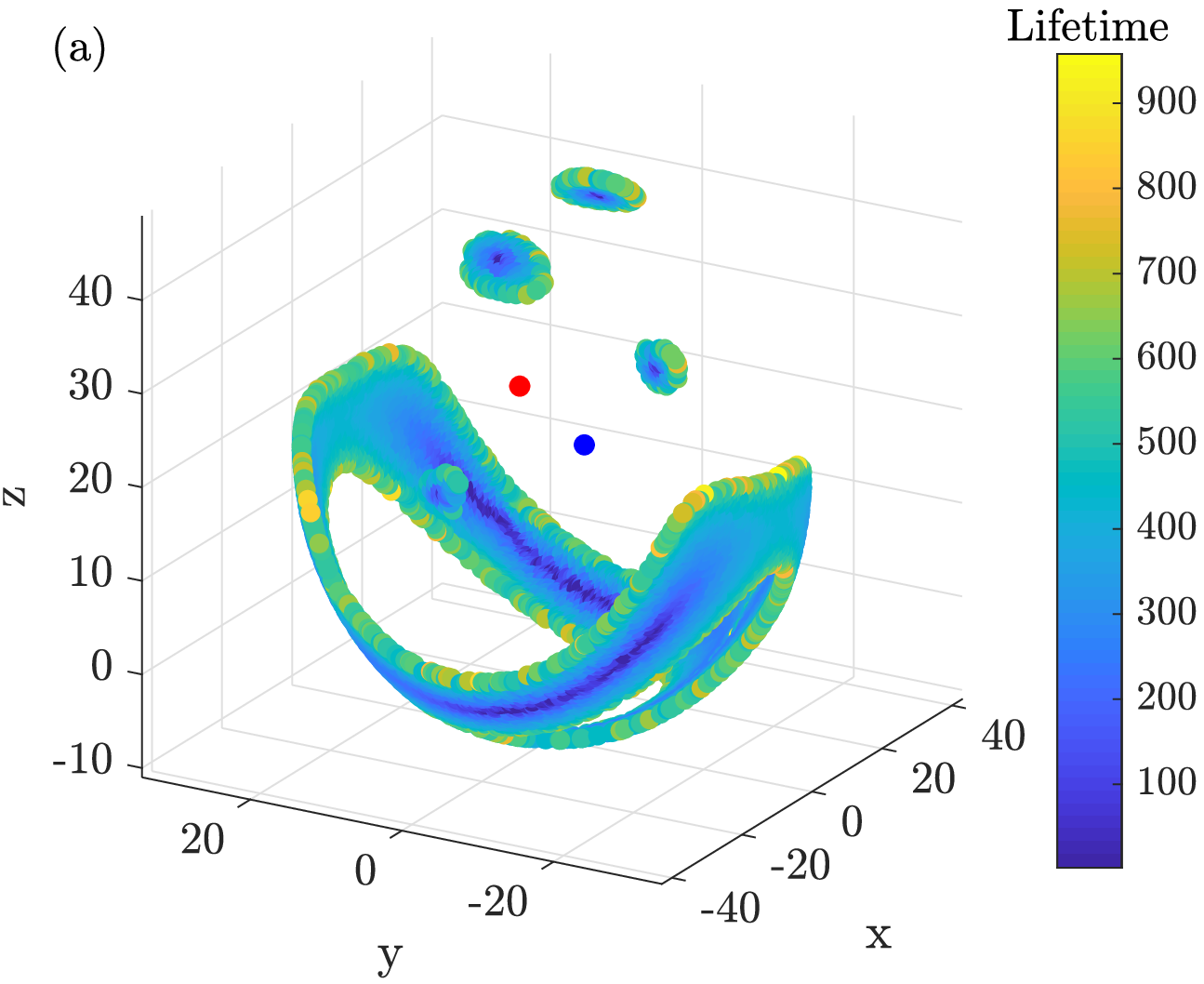}
        \includegraphics[width=0.45\textwidth]{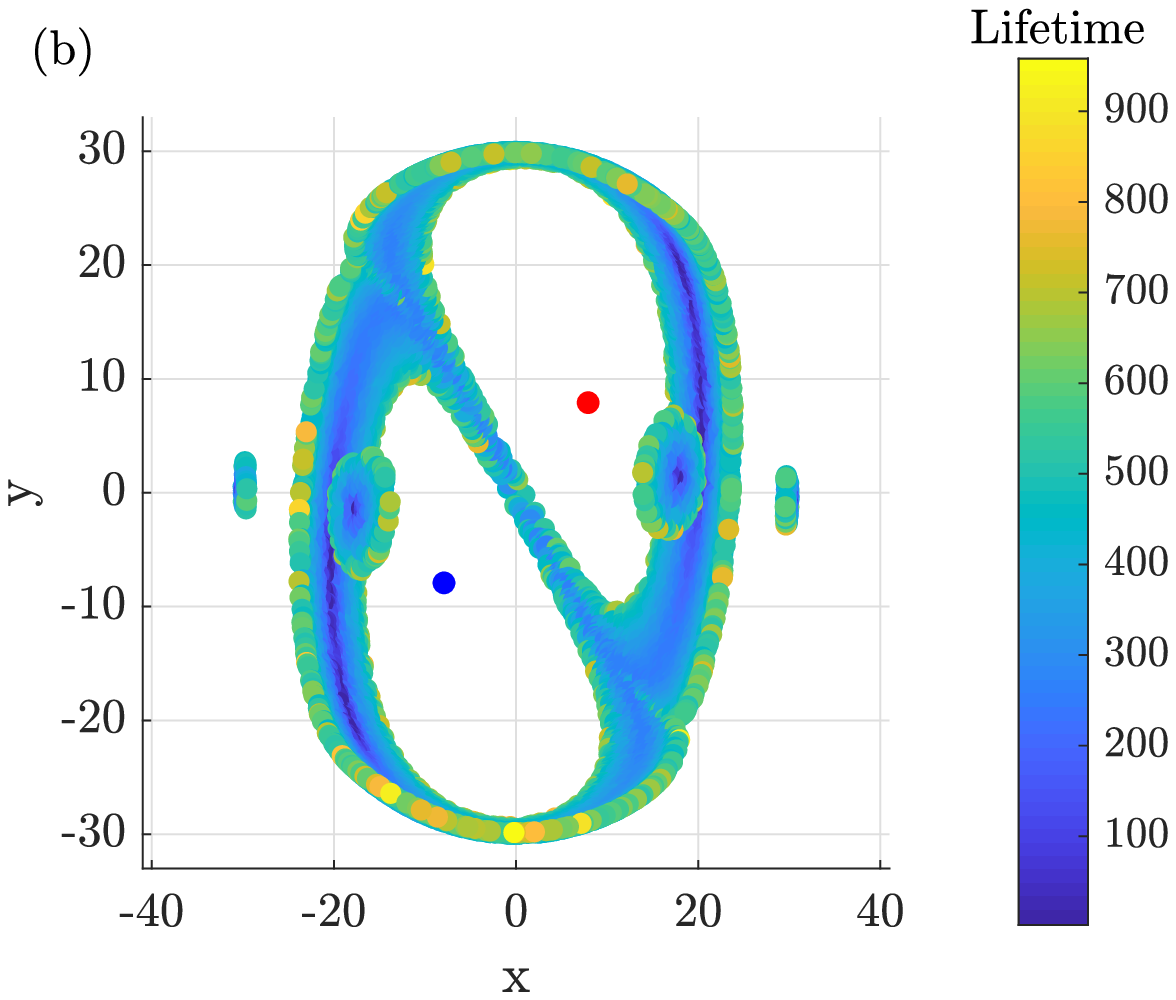}
    \end{center}
    \caption{\textbf{ Time to reach the $ C^{+}/C^{-} $ attractors for $ r=24.5 $.} (a) Red/blue points correspond to the attractors $ C^{+}/ C^{-} $. Only the initial conditions that end in $ C^{+}/C^{-} $ are depicted, the rest of the sphere corresponds to initial conditions that end up in the chaotic attractor. As we can see, the borders of the structure show higher lifetimes due to the slow oscillation decay to the $ C^{+}/C^{-} $ attractors.(b) This panel shows the $(x,y)$ plane of the previous panel.}
    \label{lifetime_r245}
\end{figure}

The lifetime of the trajectories that end up in either $ C^{+}/C^{-} $ were also computed.  We can observe that there is a wide range of lifetimes in Fig.~\ref{lifetime_r245}. This is not due to transient chaos but to the fact that the oscillations around  $ C^{+}/C^{-} $ are slowly damped until they reach the attractor.

\section{Dynamic Heteroclinic Bifurcation} \label{dynamic hetoclinic}

In this section, we explore the dynamics of the system when the parameter that accounts for the temperature gradient, $ r $, slowly varies. The bifurcation diagram in Fig.~\ref{Bifurcation_diagram} represents the dynamics for each value of $ r $ when the system
evolves with the corresponding fixed value of $ r $. We are interested in the case when the parameter varies during the evolution of the system. In other words, the parameter turns into a
slowly varying function of time of the form: $ r=r_{0}+\varepsilon t $, where $ \varepsilon $ is a sufficiently small parameter compared to the natural time scale of the system. Bifurcations that are crossed due to this parameter time-dependence are called dynamic bifurcations \cite{Benoit1991}.

As previously stated, this type of bifurcations have been studied when they imply the appearance/disappearance of regular attractors. For example, the dynamic pitchfork bifurcation for the Lorenz system has been deeply studied. It was found that when the temperature gradient is increased, convection rolls appear suddenly at a $ r>1 $, which is what is called the delay effect. When the temperature gradient is decreased they slowly decelerate and finally disappear for a $ r<1 $, showing hysteresis. Furthermore, they always follow the same equilibrium, i.e., roll in the same direction, to which one depends on the initial conditions. The area enclosed in the hysteresis diagram depends on the adiabatic parameter, $ \varepsilon $, and this relation is defined by its corresponding scaling law \cite{Berglund1999}.

As far as we know, the crossing of a bifurcation due to a slow parameter drift when chaotic attractors are implied has been only studied so far for maps, for instance, in the Lorenz map in \cite{Maslennikov2013} and \cite{Maslennikov2018}. Here, we study the dynamic heteroclinic bifurcation at $ r=24.06 $ for the Lorenz system. This implies a change in the number of the attractors and the appearance/disappearance of a chaotic attractor.

In analogy with the delay effect found for the pitchfork bifurcation where the origin becomes unstable but the system tracks that path for a period of time, we start our analysis for a value of $ r>24.06 $  and we decrease this value past the heteroclinic bifurcation where the chaotic attractor is no longer stable.

The first difference in this analysis with the one for regular attractors is that single trajectories are no longer representative and do not contain all the possible dynamics, thus we follow an ensemble of trajectories starting on the same sphere of initial conditions used in the previous section.

Moreover, we refer to Eqs.~\ref{Lorenz} as the frozen-in system when $ r $ is a fixed parameter and the non-autonomous system to the same set of equations when $ r $ is a function of time in the following form
\begin{equation}
r = \threepartdef { r_{0} } {t<t_{1}} {r_{0}-\varepsilon \cdot (t-t_{1})} { t_{1}<t<t{2}} {r_{0}-\varepsilon \cdot (t_{2}-t_{1})} {t>t_{2}}
\quad \quad
 \begin{tikzpicture}[baseline]
\draw[ultra thick] (0,1) -- (-1.5,1)  node[anchor=east] {$ r_{0} $};
\draw[ultra thick] (0,1) -- (1.5,-1);
\draw[ultra thick] (1.5,-1) -- (3,-1);
\draw[ dotted] (0,1.5) -- (0,-1.5)  node[anchor=north] {$ t_{1} $};
\draw[ dotted] (1.5,1.5) -- (1.5,-1.5) node[anchor=north] {$ t_{2} $};
\end{tikzpicture}
\label{r}
\end{equation}
where $ t_{1} $ is the time for which the parameter shift starts and $ t_{2} $ when it ends. This way, we let the system evolve to its steady state, the chaotic attractor in our case, before the shift in $ r $ starts. Besides, we call $ t_{f}>t_{2}$ the final observation time.

\begin{figure}[h]
    \begin{center}
        \includegraphics[width=0.65\textwidth ]{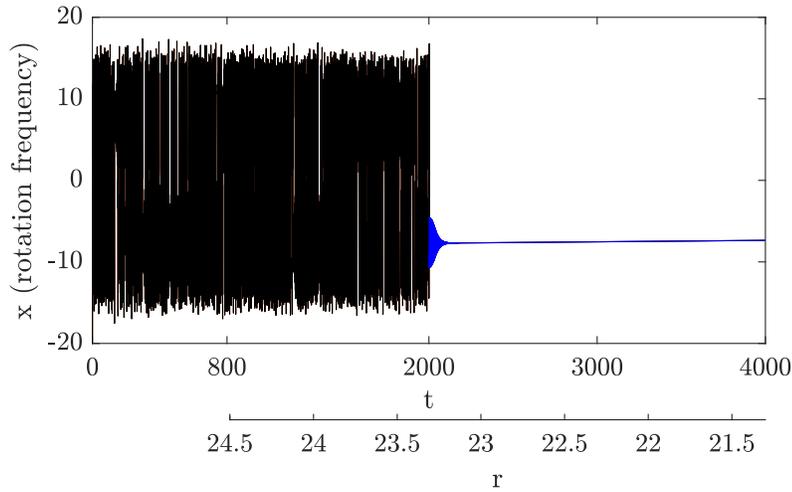}
    \end{center}
    \caption{\textbf{ Time series for the non-autonomous Lorenz system with $ \varepsilon=10^{-3} $.} The secondary $x$-axis shows the time dependence in the $ r $ parameter. It can be seen that the transient lasts for a long period of time and that the transition to the steady state starts at $ r<24.06 $, specifically, for $ r=23.3 $ (when the line turns blue). }
    \label{time_series_epsilon103}
\end{figure}

In Fig.~\ref{time_series_epsilon103}, the $x$-component of a trajectory starting at the sphere, with the following parameters: $ r_{0}=24.5 $, $ t_{1}=800 $, $ t_{2}=4000 $, $ t_{f}=4000$ and $
\varepsilon=10^{-3} $ is shown. As we can see, the motion is chaotic up until a value (when the line turns blue) for which there is a sudden transition to the attractor $ C^{-} $. Equation~\ref{r} establishes a correspondence between time and the value of the temperature gradient, $ r $, which is also shown in the figure as a secondary $x$-axis. The transition from one regime to another appears for a value of $ r<24.06 $ that we call the critical value and designate it by $ r_{cr} $.

Thus, we conclude that the delay effect is also present when chaotic attractors are involved. Furthermore, it is necessary a lower temperature gradient, exactly $ r_{cr}=23.3 $  for the previous example, to achieve a constant rotation frequency of the rolls. However, if we let the system evolve to its steady state for each value of $ r $, which translated to our analysis would mean to make $ \varepsilon \rightarrow 0$, the transition would take place exactly at the value in the bifurcation diagram, $ r=24.06 $. For $ 23.3<r<24.06 $, the system tracks the chaotic attractor although it is no longer stable. In other words, the chaotic attractor is a metastable state.

As previously mentioned, when dealing with chaotic attractors a single trajectory is not informative enough and we should consider an ensemble of trajectories. For that reason, we take $ 10^{4} $ initial conditions on the sphere, and we exclude the ones that for $ r=24.5 $ go to $ C^{+}/C^{-} $. These correspond to the red/blue basins in Fig.~\ref{basin_r245}.

The attractor toward which the trajectory goes depends on the initial conditions. But for a single initial condition, it depends on the precise moment that the trajectory is caught wandering in the chaotic attractor, that is on $ t_{1} $. Using the same terminology as in \cite{Kaszas2019}, we call the basins for the non-autonomous system: scenario-dependent basins. After this
clarification, we fix $ t_{1}=800, $ $ \varepsilon=10^{-3} $, $ r_{0}=24.5 $, $ t_{2}=4000 $ and $ t_{f}=4000$, as before and we let the ensemble of trajectories evolve.

\begin{figure}[h]
    \begin{center}
        \includegraphics[width=0.45\textwidth ]{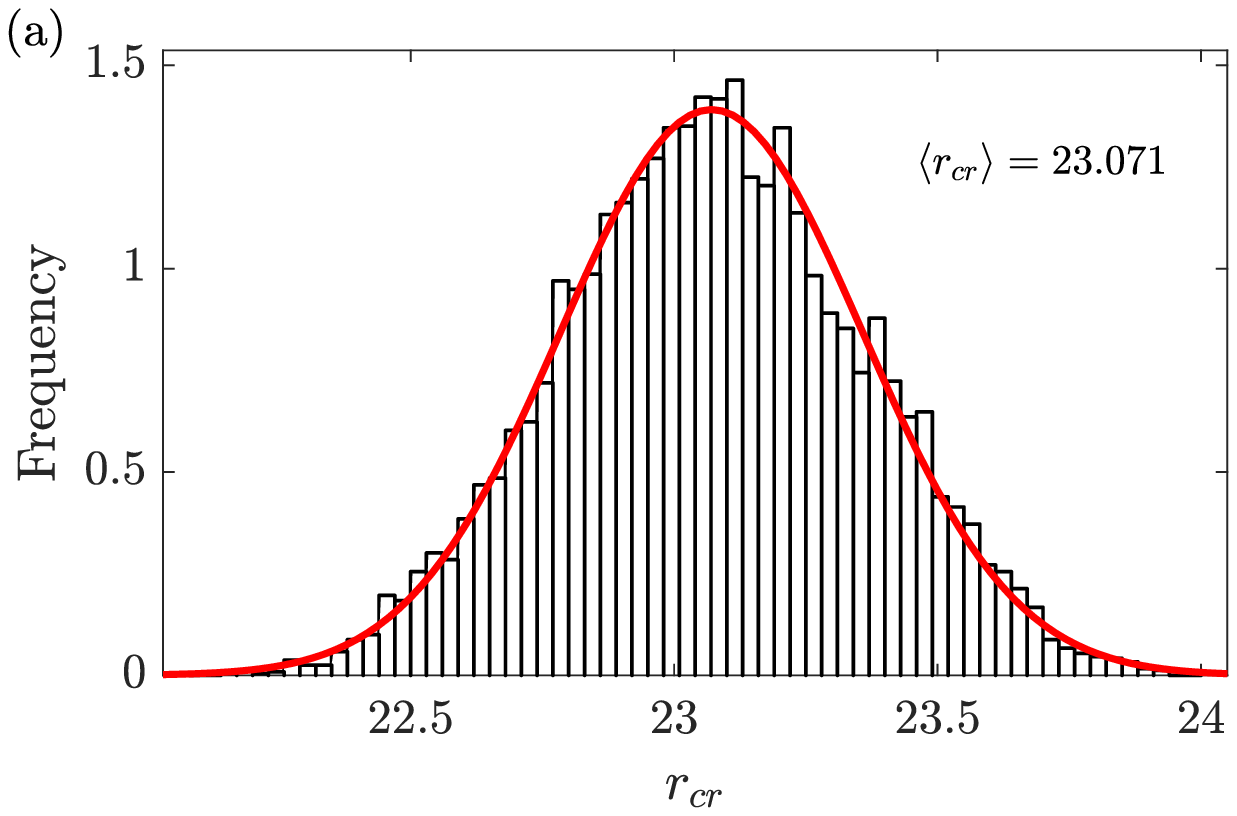}
        \includegraphics[width=0.45\textwidth ]{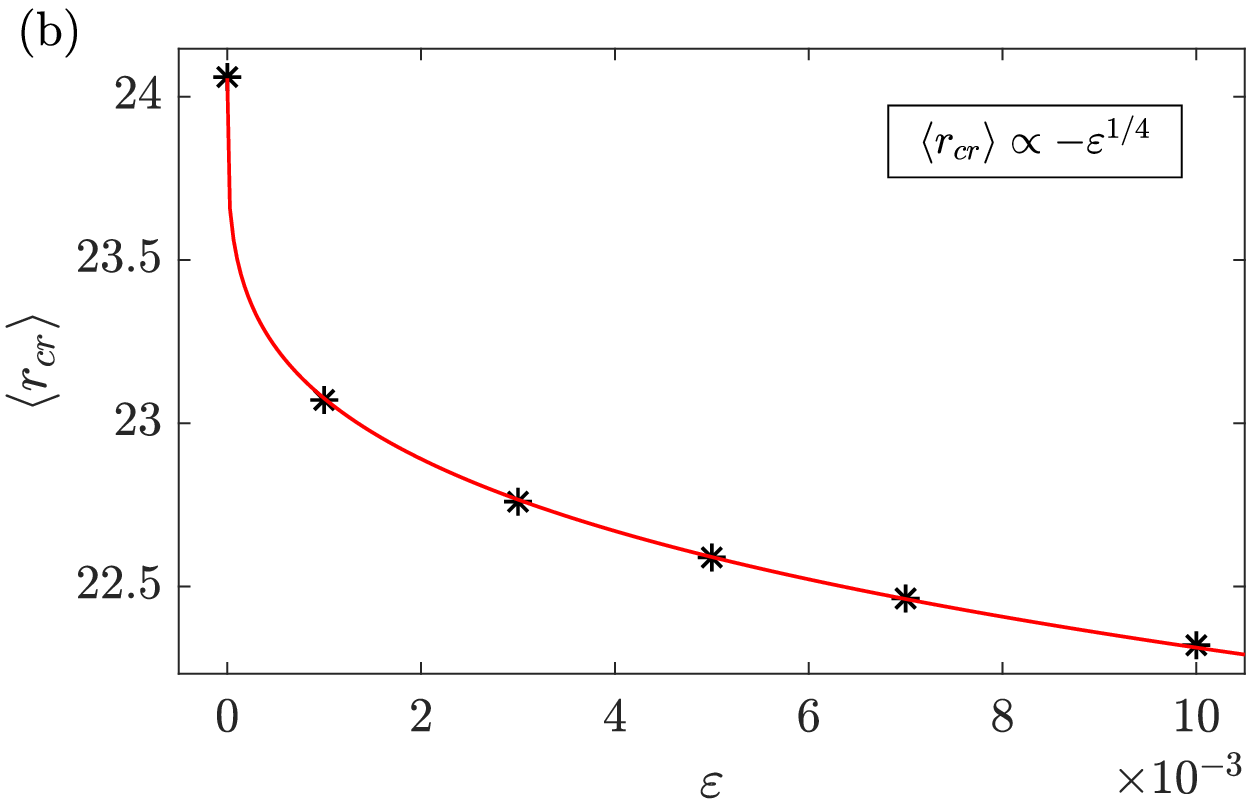}
    \end{center}
    \caption{\textbf{Critical value of the parameter $ r $ and scaling law for the heteroclinic dynamic bifurcation.} (a) The normal distribution of $ r_{cr} $ for an ensemble of trajectories with a specific rate of change of $ \varepsilon=10^{-3} $. (b) The mean value of $ r_{cr} $ for different values of $ \varepsilon $. The stars correspond to numerically calculated values of $ \langle r_{cr} \rangle $, the power law fit is also shown in red.}
    \label{histogram}
\end{figure}

The value of $ r_{cr} $ for the dynamic heteroclinic bifurcation is different for each initial condition unlike for the dynamic pitchfork bifurcation for which the transition occurred at the same value of $ r $  for every initial condition. In Fig.~\ref{histogram}(a), we can see how the values of $ r_{cr} $ follow a normal distribution with mean $ \langle r_{cr} \rangle =23.071 $ and
standard deviation $ \sigma=0.2868 $.

We may repeat the same procedure for different values of $ \varepsilon $ to study the effect of the parameter rate of change. We focus on small, but non-negligible rates compared to the natural time scale of the system. In our case, the period of revolution of convection rolls is of order $ 10^{-1} $. In Fig.~\ref{histogram}(b), we calculated the $ r_{cr} $ for more values of
$ \varepsilon $ on the range $ 10^{-4} -10^{-2} $ which are marked as black stars. As we can see, for higher rates the delay effect is more pronounced. For example, for $ \varepsilon=10^{-2} $, the
temperature gradient for which the system suffers a sudden transition from a chaotic state to a stable state is $ \langle r_{cr} \rangle=22.32  $, which is significantly a smaller value than
the $ r_{cr} $ for $ \varepsilon=10^{-3} $. Another consequence of this, is that for rapid shifts in the parameter, the metastable state lasts longer. Also, as $ \varepsilon \rightarrow 0$, the
standard deviation is reduced. In the limit, that is, the frozen-in system, all the initial conditions suffer the transition at the same value of $ \langle r_{cr} \rangle $.

Finally, we derived the corresponding scaling law for the heteroclinic bifurcation. This equation relates the value of the parameter for which the system abandons the chaotic attractor with the parameter change rate. For that purpose, points in Fig.~\ref{histogram}(b) are fitted to a power law of the form
\begin{equation}
\langle r_{cr} \rangle = -a \cdot \varepsilon^{1/4} +r_{0},
\label{scaling_law}
\end{equation}
where $ a>0 $ is a constant and $ r_{0}=24.5 $ in our case, with a $R$-square: $ R^{2}=0.9999 $. This law indicates that a small parameter change rate reduces the parameter value for the transition, but from a certain value an increase in the parameter change rate does not reduce significantly the parameter value.

We may compare this with the same analysis for the pitchfork bifurcation for which $ r_{cr} \propto \varepsilon^{1/2} $. Unlike the scaling law for the pitchfork bifurcation, our scaling law deals with the presence of chaotic attractors, which to the best of our knowledge is a novel finding. This implies that the value of $ r_{cr} $ in the scaling law is a mean value from the ensemble approach. 

We may ask ourselves about the spatial distribution of $ r_{cr} $ in the sphere of initial conditions. However, due to the presence of the chaotic attractor, there is no pattern and the initial conditions that lead to longer and shorter metastable states are completely intermingled. We may say that the chaotic attractor acts as a memory-loss agent. In the same way, the scenario-dependent basins do not show a pattern and both basins ($ C^{+}/C^{-}$: red/blue basin) are intermingled. The final destination of the trajectories is related to the precise moment that the trajectory is caught wandering in the chaotic attractor when the parameter shift starts rather than to the initial condition. In other words, predictability of individual trajectories is lost because the passage through the chaotic attractor induces fractal basins of attraction. Similar phenomena has been addressed as a random tipping in \cite{Kaszas2019}.

\subsection{Transient chaos interpretation}

So far, we have defined the delay effect for the dynamic heteroclinic bifurcation, but this phenomenon can be interpreted from a different point of view in the context of transient dynamics.
The Lorenz equations with $ r $ defined by Eq.~\ref{r} form a non-autonomous system, which can be considered to present transient chaos as the duration of the chaotic dynamics is finite (see Fig.~\ref{time_series_epsilon103}). In this context, the scaling law predicts the end of the transient state. 

In the previous section, the scaling law predicted the value of the temperature gradient for which the dynamics changed as past $ r=24.06 $ the chaotic attractor was considered to be a metastable state. In the time framework, the scaling law predicts when the system suffers a transition to its steady state. The transient dynamics may last for long periods of time; therefore, the transient is not a negligible part of the dynamics as it has been considered before, and it is fundamental to uncover a law that predicts the end of this state.

\begin{figure}[h]
    \begin{center}
        \includegraphics[width=0.65\textwidth ]{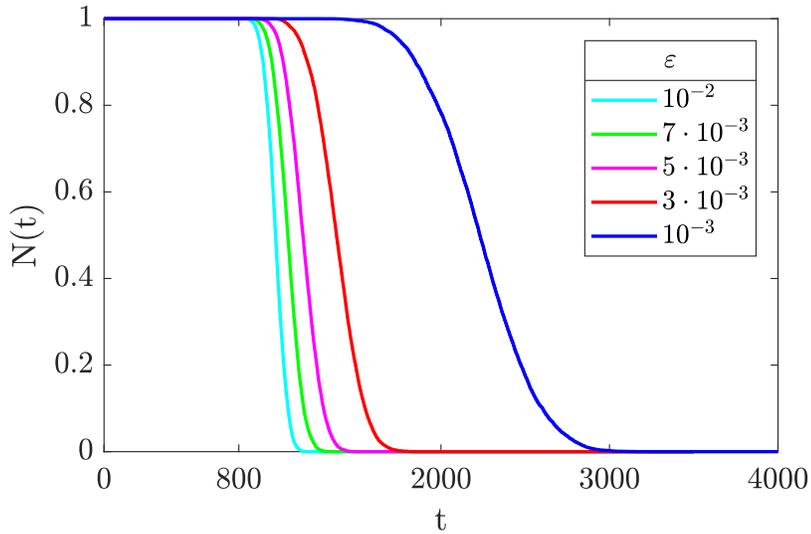}
    \end{center}
    \caption{\textbf{ Normalized number of trajectories that remain in the chaotic attractor for a time $t$.} At $ t=800 $, the parameter starts decreasing, but the chaotic attractor constitutes a metastable state for some time later than before reaching the steady state. Therefore, the transient behavior is shown. For a certain parameter change rate, the trajectories do not reach the steady state at once; instead they follow a normal distribution. For slow parameter change rates the transient dynamics lasts for a longer period of time. This may create the false impression that the transient regime is the steady state. }
    \label{decayrate}
\end{figure}

To characterize a nonattracting chaotic set, as the chaotic saddle responsible for the transient chaos, we may analyze the decay in the number of trajectories that still present a chaotic behavior \cite{Maslennikov2013}. In Fig.~\ref{decayrate} we represent, for different decay rates, $ N(t) $ as the normalized number of trajectories in the chaotic attractor for a time $ t $. As we did earlier, we exclude the ones that at $ r = 24.5 $ belong to the $ C^{+}/C^{-} $ basins. Note that the decay in $ r $ starts at $ t=800 $. When the curves decrease to zero, the transient chaos phase ends and every trajectory reaches its steady state, i.e., the $ C^{+}/C^{-} $ fixed point attractors.

As we can see, in all cases, the decay with time follows a sigmoid which is related to the normal distribution for $ r_{cr} $ in Fig.~\ref{histogram}(a). Note that $ r_{cr} $ and time are equivalent through Eq.~\ref{r}. In fact, we are representing nothing more than the complementary cumulative distribution function of Fig.~\ref{histogram}(a) in terms of time.

The $S$-shape of Fig.~\ref{decayrate} reflects that the majority of trajectories decay more or less at the same time (same $ r_{cr} $), while some of them decay earlier or later. The curve for $ \varepsilon=10^{-3} $ is further from the rest of them as we showed in Fig.~\ref{histogram}(b) that $\langle r_{cr} \rangle \propto -\varepsilon^{1/4}  $, thus as $ \varepsilon \rightarrow 0$, the $\langle r_{cr} \rangle  $ decreases and the time for the transition increases non-monotonically.

The scaling law on Eq.~\ref{scaling_law} predicted that for high parameter change rates the temperature gradient could be decreased to a low value before turbulence disappeared. In this new interpretation, we add that in terms of time, the transient dynamics is shorter in that
case. Finally, the scaling law can be written in terms of time using Eq.~\ref{r}:
\begin{equation}
\langle \tau \rangle = a \cdot \varepsilon^{-3/4}+t_{1},
\label{Scaling Law lifetime}
\end{equation} 
where $ \tau $ refers to the lifetime of the transient dynamics, $ t_{1} $ in our case is $ 800 $ and $ a $ is a positive constant. 

We conclude that in non-autonomous systems, the transient dynamics may present an unexpected long-term behavior. After a period of apparent equilibrium, in our case chaotic, the system suffers a transition to its truly steady state. The transient dynamics duration depends on the parameter rate of change through the scaling law. For slow rates, the transients last for a long period. This may be an undesired effect for a experimentalist as a parameter may be changing too slowly to notice and the dynamics may seem stable. Additionally, due to the presence of the chaotic attractor, the final destination of the trajectory after the transient dynamics becomes unpredictable.

\section{Reversibility} \label{reversibility}

Sometimes, the long-lasting transients that appear in systems with parameter drifts may be followed by an undesirable state. This problem has been addressed from a control theory approach, where a small perturbation may keep the system in the desired transient state \cite{Aguirre2004}. For instance, this approach has been proposed to prevent species extinction maintaining the system in the transient chaos regime \cite{Shulenburger1999}. Here, we tackle the problem from a different perspective.

For systems with parameter drifts, the parameter may be controllable and we may ask ourselves whether it is possible to reverse the dynamics by reversing the parameter to its original value. For example, we can see that in the non-autonomous Lorenz system, a chaotic transient precedes a sudden transition to a fixed point as shown in Fig.~\ref{time_series_epsilon103}. An intuitive way to avoid the latter, is to increase the temperature gradient again in order to recover the chaotic dynamics that was lost. However, if we analyze the frozen-in bifurcation diagram, the fixed point attractors $ C^{+}/C^{-} $ are stable before and after the heteroclinic bifurcation at $ r=24.06 $. This implies that it may not be so easy to avoid these states and a complete study of them would be needed.

For that purpose, in this section, we analyze what happens to the system if the temperature gradient increases with time and the heteroclinic bifurcation is crossed `from the left'. The equations for the parameter $r$ read:

\begin{equation}
r = \threepartdef { r_{0} } {t<t_{1}} {r_{0}+\varepsilon \cdot (t-t_{1})} { t_{1}<t<t{2}} {r_{0}+\varepsilon \cdot (t_{2}-t_{1})} {t>t_{2}}
\quad \quad
\begin{tikzpicture}[baseline]
\draw[ultra thick] (0,-1) -- (-1.5,-1)  node[anchor=east] {$ r_{0} $};
\draw[ultra thick] (0,-1) -- (1.5,1);
\draw[ultra thick] (1.5,1) -- (3,1);
\draw[ dotted] (0,1.5) -- (0,-1.5)  node[anchor=north] {$ t_{1} $};
\draw[ dotted] (1.5,1.5) -- (1.5,-1.5) node[anchor=north] {$ t_{2} $};
\end{tikzpicture}
\label{r-}
\end{equation}

Once again, we take a sphere of $ 10^5 $ initial conditions and we let these trajectories evolve with time. Regarding the parameters in Eq.~\ref{r-}, there are some restrictions in order to reverse the dynamics. First of all, we take the initial temperature gradient to be in the range $ 13.926<r_{0}<24.06 $. In particular, we consider $ r_{0}=20 $ since for this parameter we have already calculated the corresponding basins of attraction in the frozen-in system (see Fig.~\ref{basin_r20}). We aim to increase the temperature gradient in such a way that those red and blue basins corresponding to $ C^{+}/C^{-} $ (the undesirable state), map to the yellow basin corresponding to the chaotic attractor (the desirable state).

The ideal situation would be to fix $ t_{1} $ to a higher value than the maximum lifetime calculated in Fig.~\ref{lifetime_r20}, that is, $ t_{1}>140 $. By doing so, we ensure that all  initial conditions have reached $ C^{+}/C^{-} $. However, it can be checked that once a
trajectory is close enough to $ C^{+}/C^{-} $ it cannot escape, no matter the rest of the parameter values in Eq.~\ref{r-}. By close enough, we refer to the same criterion used for the basins of attraction, that is, that the trajectories enter into a sphere of radius $0.1$ centered at $ C^{+}/C^{-} $. Higher values of the radii, around $5$, mark the no return limit. Taking that argument into account, we set $ t_{1}=0 $, so that the parameter starts increasing from the beginning of the trajectory, when it is still on the sphere, far enough from the $ C^{+}/C^{-} $ attractors.

For the same reason as before, if we take slow parameter rates, the trajectories will enter the no return limit. Thus, the third restriction is about $ \varepsilon $. For the range of $ \varepsilon $ considered in the previous section, the system does not tip to the chaotic attractor. In fact, we found that a minimum rate close to $ 5 \cdot 10^{-2} $ is needed. This type of tipping, where the system fails to track a continuously changing quasi-static attractor for a certain critical rate is called rate-induced tipping or $R$-tipping \cite{Ashwin2012}. In our case, as we are not letting the system to reach $ C^{+}/C^{-} $, no traditional rate-induced tipping is possible \cite{Kaszas2019}.

On the other hand, the last parameter, $ t_{2} $, is fixed in such a way that for every $ \varepsilon $, the temperature gradient stops at $ r=24.5 $. For example, for $ \varepsilon=10^{-1} $, the value of $ t_{2} $ is 45, since $ 20+10^{-1} \cdot (45-0)=24.5 $.

\begin{figure}[h]
    \begin{center}
        \includegraphics[width=0.3\textwidth ]{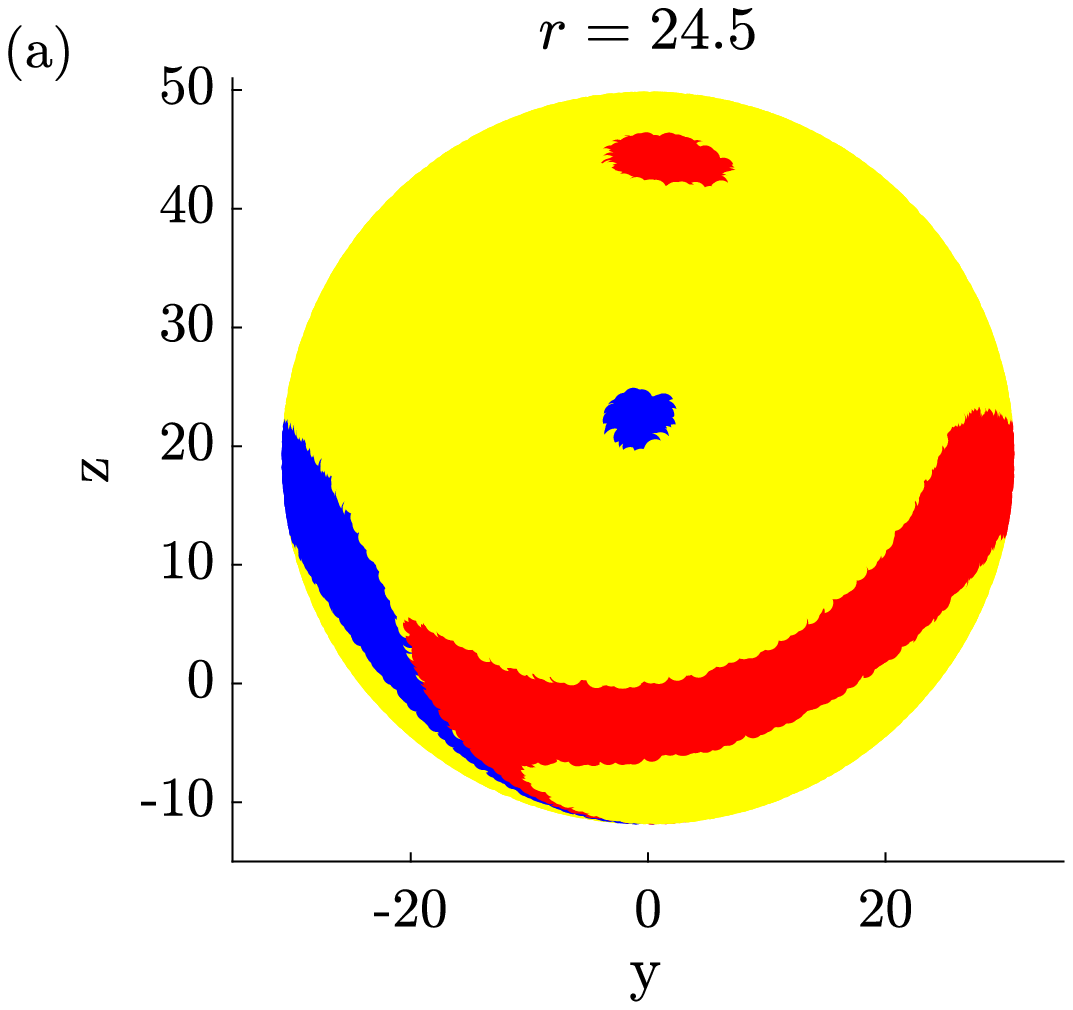}
        \includegraphics[width=0.3\textwidth ]{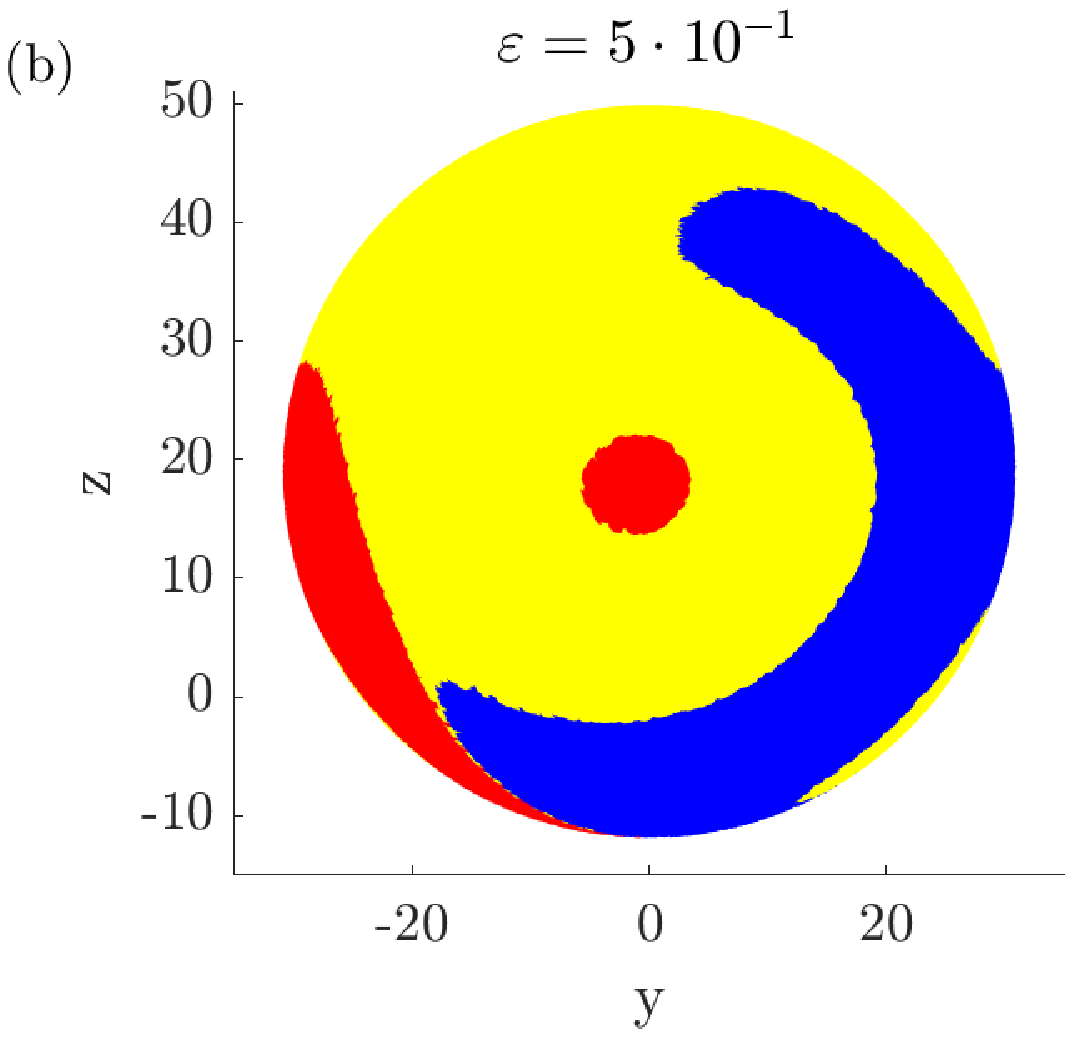}
        \includegraphics[width=0.3\textwidth ]{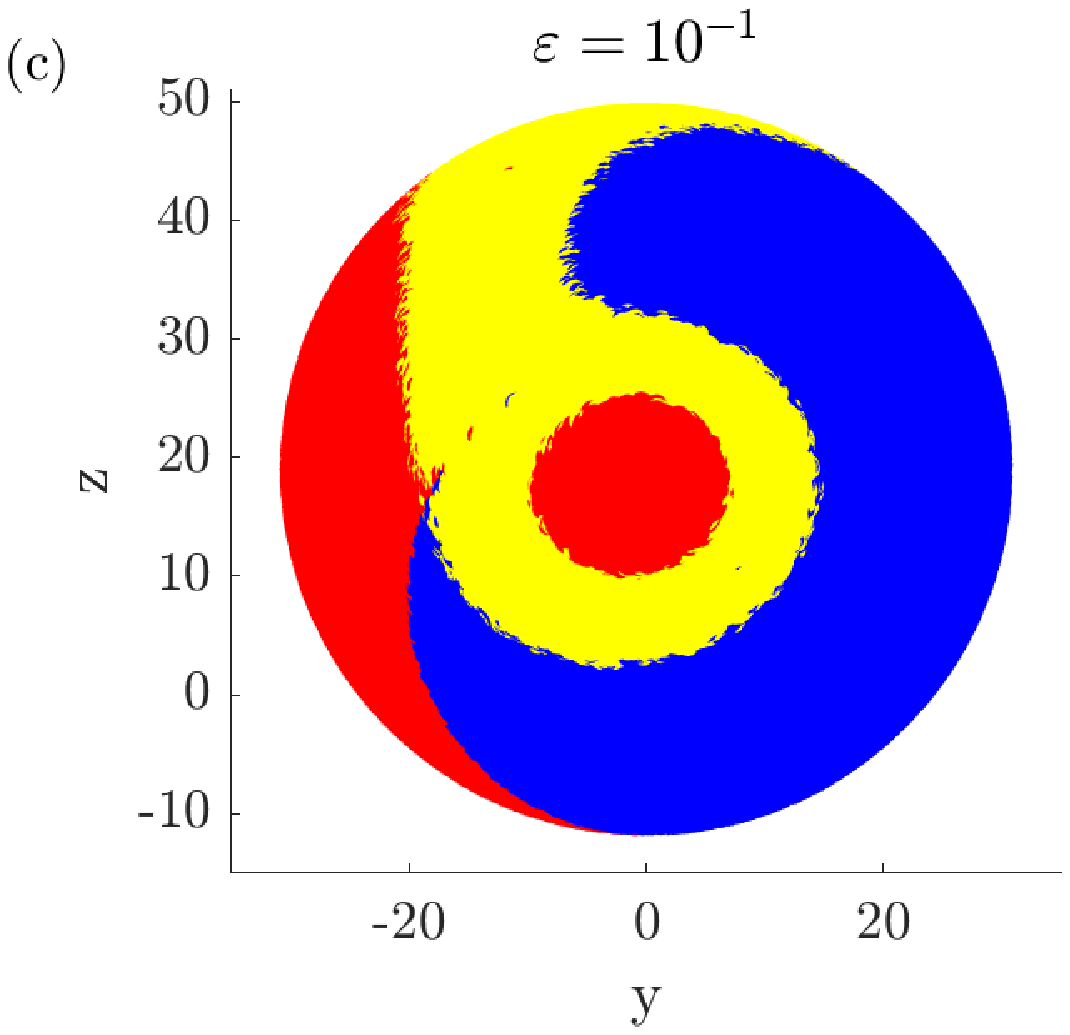}
        \includegraphics[width=0.3\textwidth ]{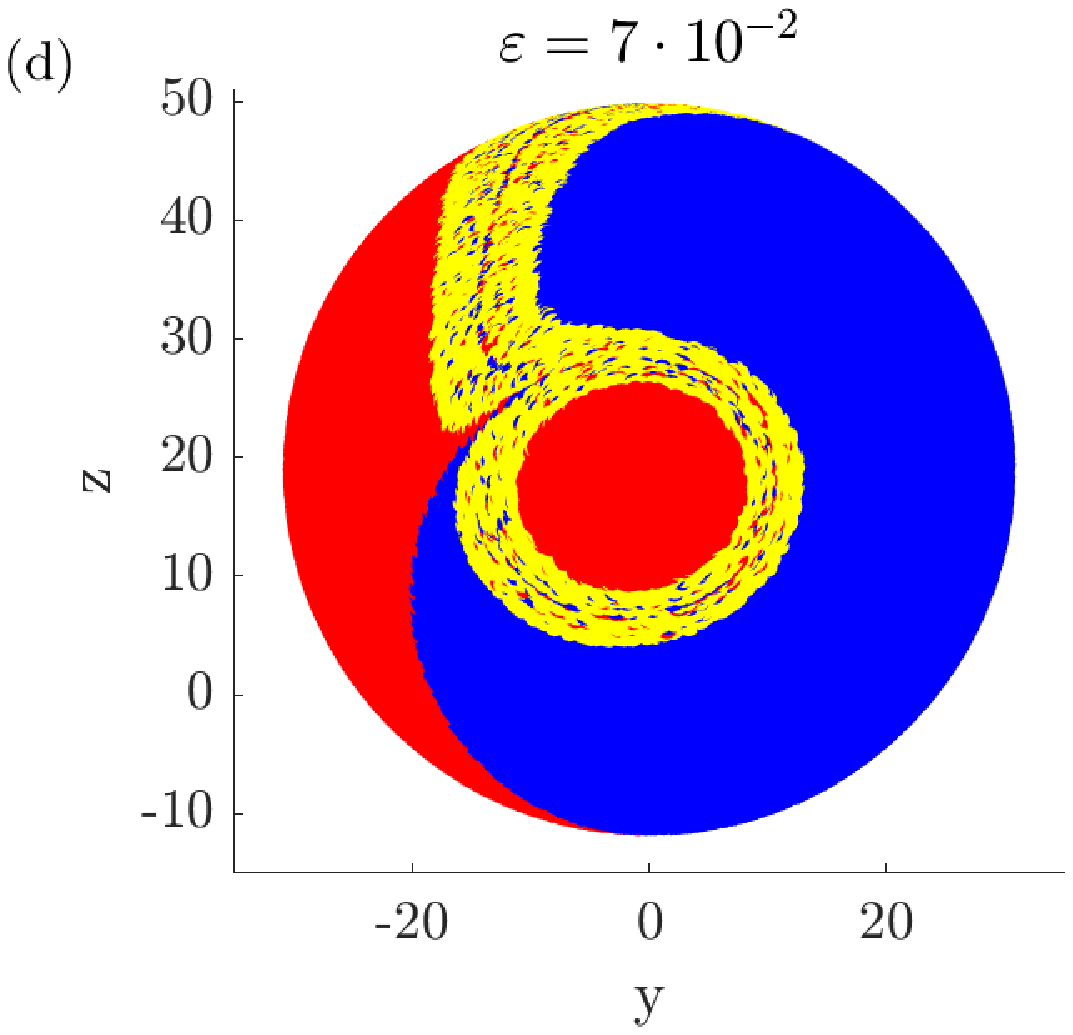}
        \includegraphics[width=0.3\textwidth ]{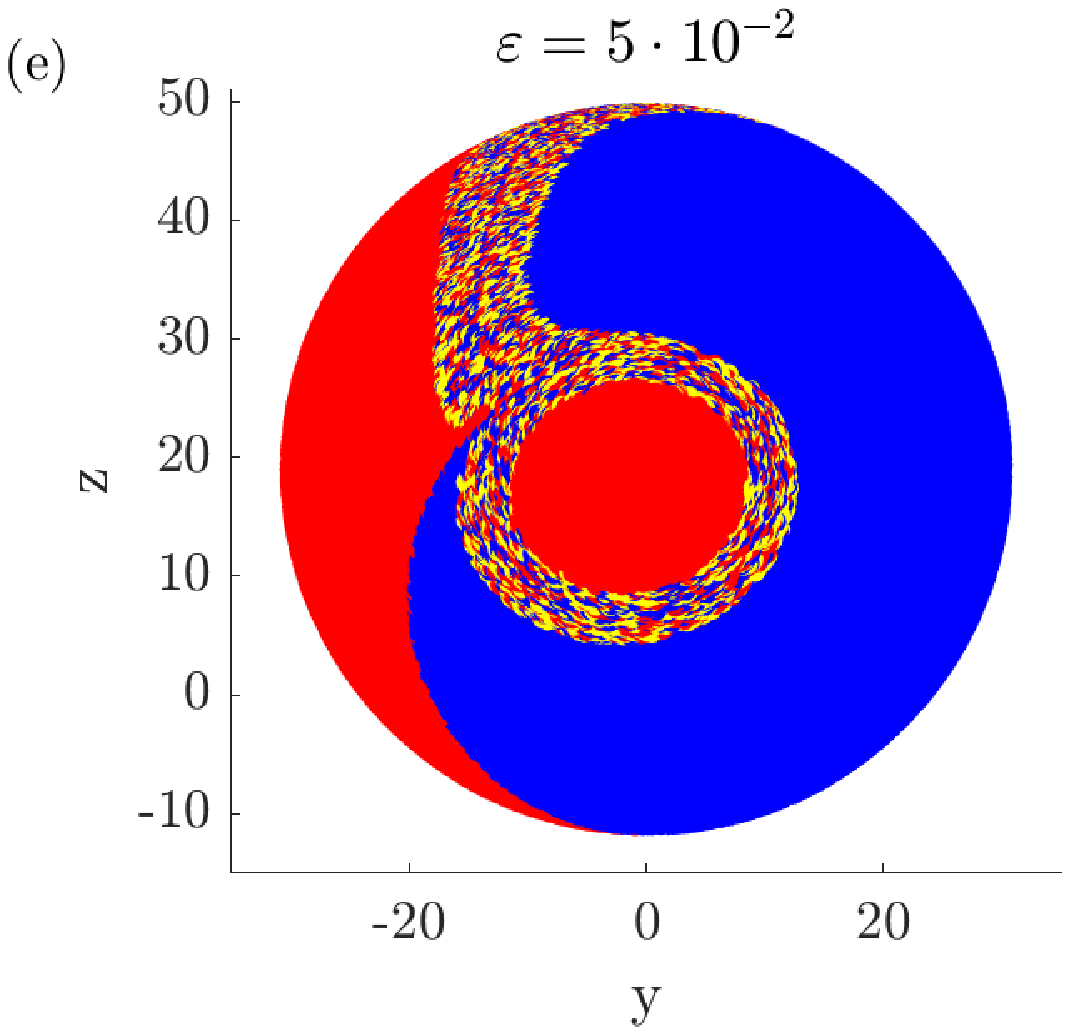}
        \includegraphics[width=0.3\textwidth ]{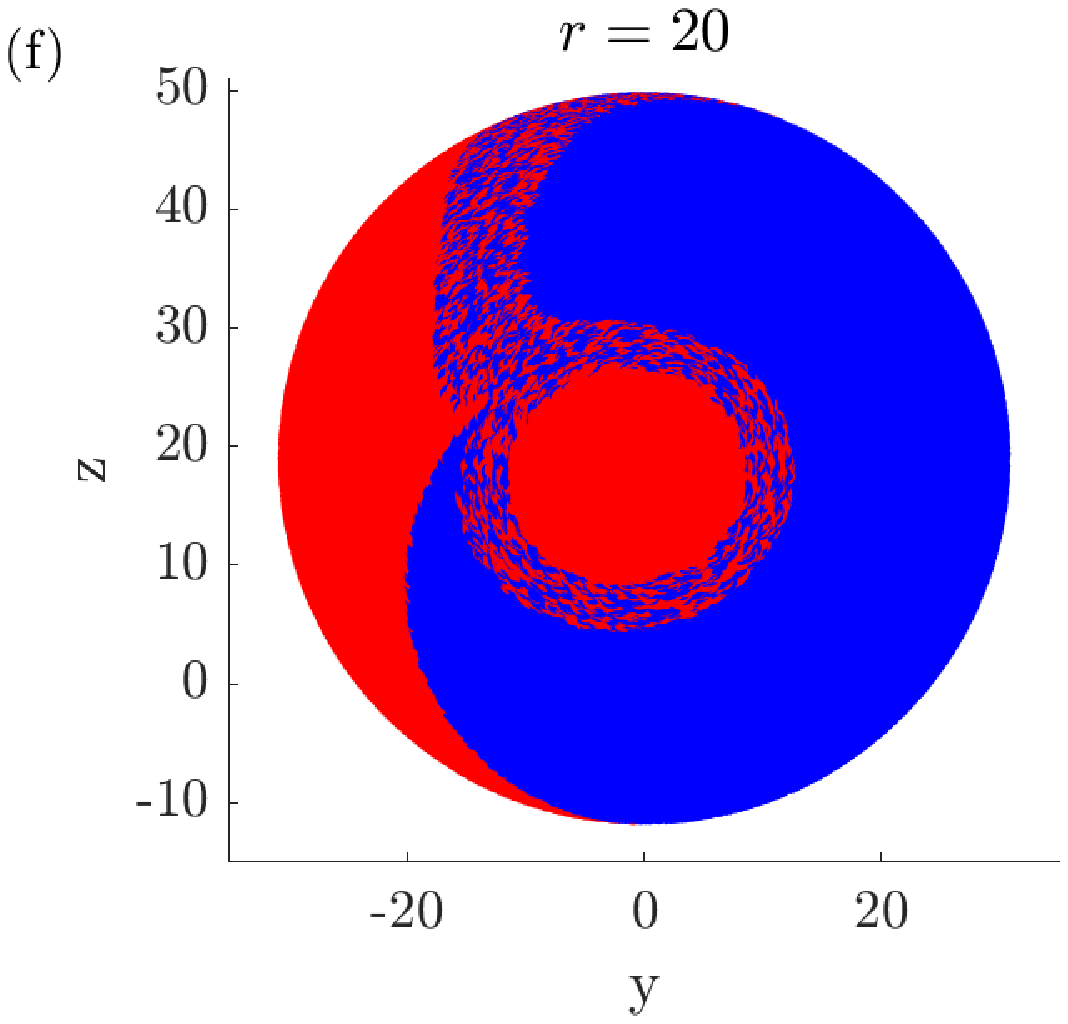}
    \end{center}
    \caption{\textbf{Rate-induced tipping phenomenon.} The scenario-dependent basins of attraction for different values of $ \varepsilon $ showing that for high parameter change rates a significant number of trajectories tip to the chaotic attractor. We also show the basins of attraction of the frozen-in system for $ r=20 $ and $ r=24.5 $ for comparison. In all cases we have fixed $ t_{1}=0, r_{0}=20 $ and $ r_{f}=24.5 $.  }
    \label{nonauto_basins}
\end{figure}

Now, we show the results for various parameter rates above the limit. Figure~\ref{nonauto_basins} shows the scenario-dependent basins for $ \varepsilon=5 \cdot 10^{-1}, 10^{-1}, 7 \cdot 10^{-2}$
and $5 \cdot 10^{-2} $. Additionally, we have included the basins for the frozen-in equations at the starting and ending values of the parameter shift, that is, $ r=20$ and $ 24.5 $, in order to compare the size of the yellow regions, corresponding to initial conditions that end up in the chaotic attractor.

As we can see, the faster the parameter shift, the closer to the basin for $ r=24.5 $. For slower parameter rates, as $ \varepsilon=10^{-2} $, we obtain that the scenario-dependent basin is the same as the basin for $ r=20 $. In other words, no tipping from $ C^{+}/C^{-} $ to the chaotic attractor is possible. For $ \varepsilon=5 \cdot 10^{-2} $, a few trajectories starting near the
stable manifold of the chaotic saddle at $ r=20 $ tip to the chaotic attractor. This is because points in that region have longer lifetimes and show transient chaos in the static case, see Fig.
~\ref{lifetime_r20}. For $ \varepsilon= 7 \cdot 10^{-2} $ and $10^{-1} $, this effect is stressed. And for $ \varepsilon=5 \cdot 10^{-1} $ the number of trajectories that tip to the chaotic
attractor are significantly increased and the basin boundaries are smoothed.

\begin{figure}[h]
    \begin{center}
        \includegraphics[width=0.65\textwidth ]{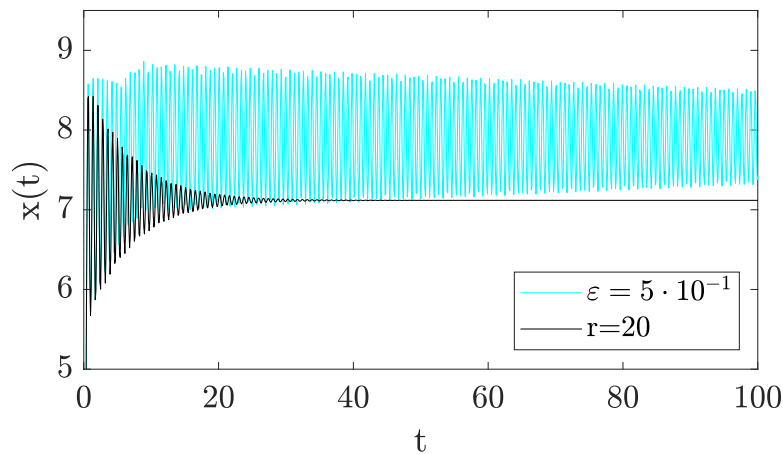}
    \end{center}
    \caption{\textbf{ Transient lifetime comparison for the frozen-in case and the non-autonomous system with $ \varepsilon=5 \cdot 10^{-1} $.} Even for the trajectories that do not tip to the chaotic attractor, the transient regime duration is increased when the parameter shifts as it is more difficult to track the attractor. }
    \label{longtransient}
\end{figure}

Even for the initial conditions that do not tip, these trajectories are affected by an increase in the time to reach the $ C^{+}/C^{-} $ attractors. This long-lasting transient is explained by the fact that the attractor is continuously changing and the difficulty to track its path is more severe for higher parameter change rates as shown in Fig.~\ref{longtransient} for $ \varepsilon= 5 \cdot 10^{-1} $. The transient lasts for a longer time compared to the frozen-in case, even if it does not tip to the chaotic attractor.

\begin{figure}[h]
    \begin{center}
        \includegraphics[width=0.65\textwidth ]{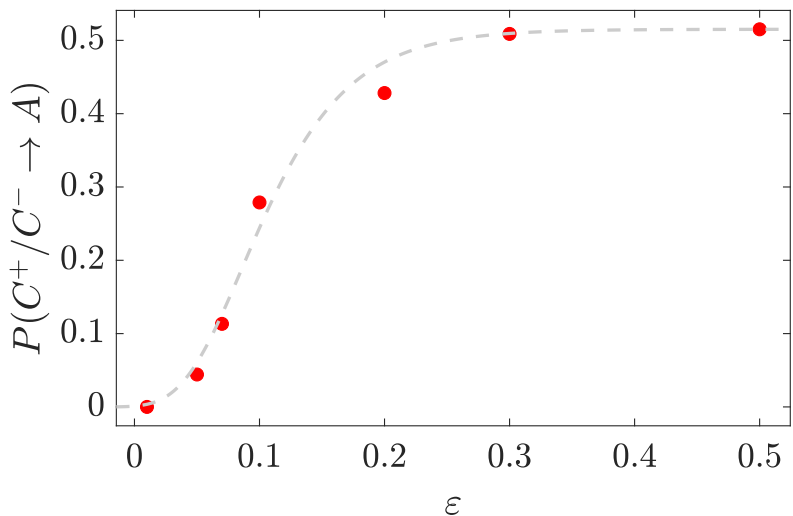}
    \end{center}
    \caption{\textbf{ Tipping probability dependence with $ \varepsilon $.} This probability accounts for the number of trajectories that tip from $ C^{+}/C^{-} $ to the chaotic attractor, $ A $, by the end of the parameter shift. For higher parameter change rates more trajectories that belonged to the red/blue basins at $ r=20 $, tip to the chaotic attractor when the parameter shifts. The numerically calculated points are fitted to a sigmoid, specifically, the Gompertz function.}
    \label{tipping_prob}
\end{figure}

Finally, we have calculated the tipping probability as defined in \cite{Kaszas2019}. In our case, it corresponds to the proportion of the part of the basin of attraction of $ C^{+}/C^{-} $
at $ r=20 $ that is mapped to the chaotic attractor, $ A $, at $ r=24.5 $ by the end of $ t_{2} $. The evolution of the tipping probability with the rate of change of the parameter $ r $ is shown
in Fig.~\ref{tipping_prob}. As we can see, the probability increases with $ \varepsilon $, since more initial conditions tip to the chaotic attractors for higher parameter change rates. At a value around $ 10^{-1} $ the curvature changes sign and around $ 5 \cdot 10^{-1} $ the tipping probability saturates and approximately half of the trajectories tip to the chaotic attractor.

The shape of the numerically calculated points describe a sigmoid, specifically, a Gompertz type function of the form
\begin{equation}
P=a \cdot e^{-b \cdot e^{-c \cdot \varepsilon}},
\end{equation}
where $ a $ is the saturation value, i.e., $ a=0.5151 $, $ b=6.139 $ and $ c $ is the growth rate. In our case, $ c=21.07 $ with an $R$-square of $ 0.9881 $. As we can see, the growth in the tipping probability is slower for low and high values of $ \varepsilon $ and it reaches a saturation value for which no matter how fast the parameter is shifted, approximately half of the trajectories never tip to the chaotic attractor. These trajectories correspond to the ones with lower lifetimes in the frozen-in case.

The results presented above show that it is possible to change the fate of trajectories that would reach the $ C^{+}/C^{-} $ attractors by increasing the parameter $ r $. At the beginning of this section, we considered a trajectory with a chaotic transient preceding a sudden transition to an undesirable state, like the one in Fig.~\ref{time_series_epsilon103}, and we asked ourselves whether it is possible to reverse this dynamics by reversing the parameter to its original value, that is $ r=24.5 $. In the light of the results shown, this is not always possible.

Once a trajectory has decayed to either one of the fixed points, that is, time has passed so that $ r $ has decreased below $ r_{cr} $, reversing the parameter to its original value is not enough to
reverse the dynamics. However, if we observe this behavior for a single trajectory, we may still be able to reverse the dynamics for other trajectories. As we showed in Fig.~\ref{histogram}(a), each
initial condition decays to the fixed point at a different time, or likewise at a different parameter value $ r $. This means that for $ r> (\langle r_{cr} \rangle - \sigma) $, there are still trajectories in the chaotic region which after some time would decay to $ C^{+}/C^{-} $, and an increase in the temperature gradient may keep these in the desirable state. For higher parameter change rates, a higher number of trajectories will tip to the chaotic attractor as shown in Fig.~\ref{tipping_prob}.

Another consequence is that if we want to make sure that the chaotic behavior is not recovered once the temperature gradient is increased for any trajectory, we have to decrease this parameter to a value of $ r<  \langle r_{cr} \rangle  - \sigma $.

\section{Conclusions}\label{conclusions}

In this paper we have studied the phenomenon of transient dynamics in a non-autonomous system. In particular, we have analyzed the Lorenz system subjected to a small parameter drift. First of all, we have characterized the associated frozen-in system. We have observed that when the parameter crossed the heteroclinic bifurcation value, the system continued tracking the chaotic attractor for further parameter values until it reached a critical value at which it jumped to its truly steady state. Thus, we have concluded that the so-called delay effect is also present in systems with strange attractors. Furthermore, it constitutes the origin for the long-term transient dynamics before the system settles at its steady state.

We derived a scaling law to relate the duration of the transient in the non-autonomous system to the parameter rate of change. We have shown that for higher parameter change rates, the transient dynamics is shorter, while the deviation from the bifurcation value of the parameter that causes the transition is larger. This relation is governed by a power law. 

Finally, we have analyzed the possibility of recovering the transient dynamics by reversing the parameter value to its original state, as an alternative to the control theory for non-autonomous systems. For this purpose, higher parameter rates were considered. Above a critical rate, rate-induced tipping takes place and some trajectories initially far from the fixed point attractors tip back to the chaotic attractor. We also have showed the sigmoid relation between the number of trajectories that change their fate and end up in the chaotic attractor and the parameter change rate. For this purpose, the reverse on the parameter must start before $ r > \langle r_{cr} \rangle - \sigma $. Even for the trajectories for which the system does not tip, we have showed that the transient dynamics duration is enlarged.

\FloatBarrier \nonumsection{Acknowledgments} \noindent This work has been supported by the Spanish State Research Agency (AEI) and the European Regional Development Fund (ERDF, EU) under Projects No.~FIS2016-76883-P and No.~PID2019-105554GB-I00.


\begin{thebibliography}{29}
\newcommand{\enquote}[1]{``#1''}
\providecommand{\natexlab}[1]{#1}
\providecommand{\url}[1]{\texttt{#1}}
\providecommand{\urlprefix}{URL }
\expandafter\ifx\csname urlstyle\endcsname\relax
  \providecommand{\doi}[1]{doi:\discretionary{}{}{}#1}\else
  \providecommand{\doi}{doi:\discretionary{}{}{}\begingroup
  \urlstyle{rm}\Url}\fi

\bibitem[{Aguirre \emph{et~al.}(2004)Aguirre, D'ovidio \&
  Sanju{\'{a}}n}]{Aguirre2004}
Aguirre, J., D'Ovidio, F. \& Sanju{\'{a}}n, M. A. F. [2004]
  \enquote{{Controlling chaotic transients: Yorke's game of survival},}
  \emph{Phys. Rev. E} \textbf{69},  16203.

\bibitem[{Ashwin \emph{et~al.}(2012)Ashwin, Wieczorek, Vitolo \&
  Cox}]{Ashwin2012}
Ashwin, P., Wieczorek, S., Vitolo, R. \& Cox, P. [2012] \enquote{{Tipping
  points in open systems: Bifurcation, noise-induced and rate-dependent
  examples in the climate system},} \emph{Philos. Trans. R. Soc. London, Ser.
  A} \textbf{370},  1166--1184.

\bibitem[{Baer \emph{et~al.}(1989)Baer, Erneux \& Rinzel}]{Baer1989}
Baer, S. M., Erneux, T. \& Rinzel, J. [1989] \enquote{{The slow passage through
  a Hopf bifurcation: Delay, memory effects, and resonance},} \emph{SIAM J.
  Appl. Math.} \textbf{49},  55--71.

\bibitem[{Beno{\^{i}}t(1991)}]{Benoit1991}
Beno{\^{i}}t, E. [1991] \emph{{Dynamic bifurcations},} {Lecture Notes
  in Mathematics, vol. 1498} (Springer, Berlin).

\bibitem[{Berglund(2000)}]{Berglund2000}
Berglund, N. [2000] \enquote{{Dynamic bifurcations: Hysteresis, scaling laws
  and feedback control},} \emph{Prog. Theor. Phys. Supplement} \textbf{139},
  325--336.

\bibitem[{Berglund \& Kunz(1999)}]{Berglund1999}
Berglund, N. \& Kunz, H. [1999] \enquote{{Memory effects and scaling laws in
  slowly driven systems},} \emph{J. Phys. A: Math. Gen.} \textbf{32},  15.

\bibitem[{Bo-Cheng \emph{et~al.}(2010)Bo-Cheng, Zhong \&
  Jian-Ping}]{Bo-Cheng2010}
Bo-Cheng, B., Zhong, L. \& Jian-Ping, X. [2010] \enquote{{Transient chaos in a smooth memristor oscillator},} \emph{Chin. Phys. B} \textbf{19},  030510.

\bibitem[{Doedel \emph{et~al.}(2006)Doedel, Krauskopf \& Osinga}]{Doedel2006}
Doedel, E. J., Krauskopf, B. \& Osinga, H. M. [2006] \enquote{{Global
  bifurcations of the Lorenz manifold},} \emph{Nonlinearity} \textbf{19},
  2947.

\bibitem[{Hastings(2004)}]{Hastings2004}
Hastings, A. [2004] \enquote{{Transients: The key to long-term ecological
  understanding?}} \emph{Trends Ecol. Evol.} \textbf{19},  39--45.

\bibitem[{Hastings \emph{et~al.}(2018)Hastings, Abbott, Cuddington, Francis,
  Gellner, Lai, Morozov, Petrovskii, Scranton \& Zeeman}]{Hastings2018}
Hastings, A., Abbott, K. C., Cuddington, K., Francis, T., Gellner, G., Lai,
  Y. C., Morozov, A., Petrovskii, S., Scranton, K. \& Zeeman, M.L. [2018]
  \enquote{{Transient phenomena in ecology},} \emph{Science} \textbf{361}.

\bibitem[{Kasz{\'{a}}s \emph{et~al.}(2019)Kasz{\'{a}}s, Feudel \&
  T{\'{e}}l}]{Kaszas2019}
Kasz{\'{a}}s, B., Feudel, U. \& T{\'{e}}l, T. [2019] \enquote{{Tipping
  phenomena in typical dynamical systems subjected to parameter drift},}
  \emph{Sci. Rep.} \textbf{9},  1--12.

\bibitem[{Lai \& T{\'{e}}l(2011)}]{Lai2011}
Lai, Y.~C. \& T{\'{e}}l, T. [2011] \emph{{Transient chaos: Complex dynamics on
finite timescales},}
   (Springer, New York).

\bibitem[{Li \emph{et~al.}(1990)Li, Win, Weiss \& Heckenberg}]{Li1990}
Li, M.Y., Win, T., Weiss, C. O. \& Heckenberg, N.~R. [1990]
  \enquote{{Attractor properties of laser dynamics: A comparison of NH3-laser
  emission with the Lorenz model},} \emph{Opt. Commun.} \textbf{80},  119--126.

\bibitem[{Lorenz(1963)}]{Lorenz1963}
Lorenz, E. N. [1963] \enquote{{Deterministic nonperiodic
  flow},} \emph{J. Atmos. Sci.} \textbf{20},  130--141.

\bibitem[{Maslennikov \& Nekorkin(2013)}]{Maslennikov2013}
Maslennikov, O. V. \& Nekorkin, V. I. [2013] \enquote{{Dynamic boundary crisis
  in the Lorenz-type map},} \emph{Chaos} \textbf{23},  023129.

\bibitem[{Maslennikov \emph{et~al.}(2018)Maslennikov, Nekorkin \&
  Kurths}]{Maslennikov2018}
Maslennikov, O. V., Nekorkin, V. I. \& Kurths, J. [2018] \enquote{{Transient
  chaos in the Lorenz-type map with periodic forcing},} \emph{Chaos}
  \textbf{28},  033107.

\bibitem[{Morozov \emph{et~al.}(2020)Morozov, Abbott, Cuddington, Francis,
  Gellner, Hastings, Lai, Petrovskii, Scranton \& Zeeman}]{Morozov2020}
Morozov, A., Abbott, K., Cuddington, K., Francis, T., Gellner, G., Hastings,
  A., Lai, Y. C., Petrovskii, S., Scranton, K. \& Zeeman, M. L. [2020]
  \enquote{{Long transients in ecology: Theory and applications},} \emph{Phys.
  Life Rev.} \textbf{32},  1--40.

\bibitem[{Morozov \emph{et~al.}(2016)Morozov, Banerjee \&
  Petrovskii}]{Morozov2016}
Morozov, A. Y., Banerjee, M. \& Petrovskii, S. V. [2016] \enquote{{Long-term
  transients and complex dynamics of a stage-structured population with time
  delay and the Allee effect},} \emph{J. Theor. Biol.} \textbf{396},  116--124.

\bibitem[{Neishtadt(1987)}]{Neishtadt1987}
Neishtadt, A. I. [1987] \enquote{{Persistence of stability loss for dynamical
  bifurcations I},} \emph{Differ. Equations} \textbf{23},  1385--1391.

\bibitem[{Neishtadt(1988)}]{Neishtadt1988}
Neishtadt, A. I. [1988] \enquote{{Persistence of stability loss for dynamical
  bifurcations II},} \emph{Differ. Equations} \textbf{24},  171--176.

\bibitem[{Picozzi \emph{et~al.}(2019)Picozzi, Bindi, Zollo, Festa \&
  Spallarossa}]{Picozzi2019}
Picozzi, M., Bindi, D., Zollo, A., Festa, G. \& Spallarossa, D. [2019]
  \enquote{{Detecting long-lasting transients of earthquake activity on a fault
  system by monitoring apparent stress, ground motion and clustering},}
  \emph{Sci. Rep.} \textbf{9},  1--11.

\bibitem[{Poland(1993)}]{Poland1993}
Poland, D. [1993] \enquote{{Cooperative catalysis and chemical chaos: A
  chemical model for the Lorenz equations},} \emph{Physica D} \textbf{65},
  86--99.

\bibitem[{Rabinovich \emph{et~al.}(2008)Rabinovich, Huerta \&
  Laurent}]{Rabinovich2008}
Rabinovich, M., Huerta, R. \& Laurent, G. [2008] \enquote{{Neuroscience:
  Transient dynamics for neural processing},} \emph{Science} \textbf{321},
  48--50.

\bibitem[{Rabinovich \emph{et~al.}(2006)Rabinovich, Varona, Selverston \&
  Abarbanel}]{Rabinovich2006}
Rabinovich, M. I., Varona, P., Selverston, A. I. \& Abarbanel, H.D. [2006]
  \enquote{{Dynamical principles in neuroscience},} \emph{Rev. Mod. Phys.}
  \textbf{78},  1213.

\bibitem[{Shulenburger \emph{et~al.}(1999)Shulenburger, Lai, Yal{\c{c}}inkaya
  \& Holt}]{Shulenburger1999}
Shulenburger, L., Lai, Y. C., Yal{\c{c}}inkaya, T. \& Holt, R.D. [1999]
  \enquote{{Controlling transient chaos to prevent species extinction},}
  \emph{Phys. Lett. A} \textbf{260},  156--161.

\bibitem[{T{\'{e}}l \emph{et~al.}(2006)T{\'{e}}l, Gruiz \& Kulacsy}]{Tel2006}
T{\'{e}}l, T. \& Gruiz, M. [2006] \emph{{Chaotic dynamics: An
  introduction based on classical mechanics}}, (Cambridge University Press, UK).

\bibitem[{Thrane \emph{et~al.}(2015)Thrane, Mandic \& Christensen}]{Thrane2015}
Thrane, E., Mandic, V. \& Christensen, N. [2015] \enquote{{Detecting very
  long-lived gravitational-wave transients lasting hours to weeks},}
  \emph{Phys. Rev. D} \textbf{91},  104021.

\bibitem[{Wang \emph{et~al.}(2016)Wang, Lai \& Grebogi}]{Wang2016}
Wang, G., Lai, Y. C. \& Grebogi, C. [2016] \enquote{{Transient chaos-a
  resolution of breakdown of quantum-classical correspondence in
  optomechanics},} \emph{Sci. Rep.} \textbf{6},  1--13.

\bibitem[{Warecki \& Gajdzica(2014)}]{Warecki2014}
Warecki, J. \& Gajdzica, M. [2014] \enquote{{Long lasting transients in power
  filter circuits},} \emph{Comput. Appl. Elect. Eng.} \textbf{12},  324--333.

\end{thebibliography}
\bibliographystyle{ws-ijbc}

\end{document}